# ROTATING DUST SOLUTIONS OF EINSTEIN'S EQUATIONS WITH 3-DIMENSIONAL SYMMETRY GROUPS PART 3: ALL KILLING FIELDS LINEARLY INDEPENDENT OF $u^\alpha$ AND $w^\alpha$

Andrzej Krasiński
N. Copernicus Astronomical Center and College of Science
Polish Academy of Sciences, Bartycka 18, 00 716 Warszawa, Poland
email: akr@alfa.camk.edu.pl

**Abstract.** This is the third and last part of a series of 3 papers. Using the same method and the same coordinates as in parts 1 and 2, rotating dust solutions of Einstein's equations are investigated that possess 3-dimensional symmetry groups, under the assumption that each of the Killing vectors is linearly independent of velocity $u^\alpha$ and rotation $w^\alpha$ at every point of the spacetime region under consideration. The Killing fields are found and the Killing equations are solved for the components of the metric tensor in every case that arises. No progress was made with the Einstein equations in any of the cases, and no previously known solutions were identified. A brief overview of literature on solutions with rotating sources is given.

## I. Summary of the method.

This paper is the third and last part of a series of 3 papers, for parts 1 and 2 see Refs. 1 and 2 (Paper 1 and Paper 2). For convenience of the readers, this section is repeated after Paper 2.

This is a concise summary of results that will be used in this paper. For proofs, motivations and references see Paper 1[1].

Every timelike vector field $u^\alpha$ of unit length that has zero acceleration and nonzero rotation defines the functions $\tau(x), \eta(x)$ and $\xi(x)$ such that:

$$u_\alpha = \tau,_\alpha + \eta\xi,_\alpha . \tag{1.1}$$

These functions are defined up to the transformations:

$$\tau = \tau' - S(\xi', \eta'), \qquad \xi = F(\xi', \eta'), \qquad \eta = G(\xi', \eta'), \tag{1.2}$$

where the functions $F$ and $G$ obey:

$$F,_{\xi'} G,_{\eta'} - F,_{\eta'} G,_{\xi'} = 1, \tag{1.3}$$

(this guarantees that the Jacobian of the transformation is 1), and $S$ is determined by:

$$S,_{\xi'} = GF,_{\xi'} - \eta', \qquad S,_{\eta'} = GF,_{\eta'} . \tag{1.4}$$

If $u^\alpha$ is the velocity field of a fluid whose number of particles is conserved:

$$(\sqrt{-g} n u^\alpha),_\alpha = 0, \tag{1.5}$$

(where $g$ is the determinant of the metric tensor and $n$ is the particle number density), then one more function $\zeta(x)$ exists such that:

$$\sqrt{-g} n u^\alpha = \varepsilon^{\alpha\beta\gamma\delta} \xi,_\beta \, \eta,_\gamma \, \zeta,_\delta \,, \tag{1.6}$$

and it is determined up to the transformations:

$$\zeta = \zeta' + T(\xi', \eta'). \tag{1.7}$$

Note that $n$ is not defined uniquely by (1.5). For example, if $u^\alpha = \delta^\alpha{}_0$ and $n$ obeys (1.5), then $n' = nf(x, y, z)$ (where $f$ is an arbitrary function) will also obey (1.5). This nonuniqueness allows for a greater freedom in the choice of $\zeta$ than (1.7), and the freedom will be used in some cases.

The following relations hold:

$$u^\alpha \tau,_\alpha = 1, \qquad u^\beta \xi,_\beta = u^\beta \eta,_\beta = u^\beta \zeta,_\beta = 0.$$

$$\frac{\partial(\tau, \eta, \xi, \zeta)}{\partial(x^0, x^1, x^2, x^3)} = \sqrt{-g} n \neq 0. \tag{1.8}$$

The last of (1.8) guarantees that $\{\tau, \xi, \eta, \zeta\}$ can be chosen as coordinates, they will be called Plebański coordinates. Then, with $\{\tau, \xi, \eta, \zeta\} = \{x^0, x^1, x^2, x^3\} = \{t, x, y, z\}$:

$$u_\alpha = \delta^\alpha{}_0, \qquad u_\alpha = \delta^0{}_\alpha + y\delta^1{}_\alpha,$$

$$g_{00} = 1, \qquad g_{01} = y, \qquad g_{02} = g_{03} = 0, \qquad g = \det(g_{\alpha\beta}) = -n^{-2},$$

$$w^\alpha = n\delta^\alpha_3, \qquad \omega_{\alpha\beta} = -\omega_{\beta\alpha} = (1/2)\delta^1{}_\alpha \delta^2{}_\beta, \tag{1.9}$$

where $w^\alpha$ is the rotation vector field, and $\omega_{\alpha\beta}$ is the rotation tensor corresponding to the velocity field $u_\alpha$:

$$\omega_{\alpha\beta} = \frac{1}{2}(u_{\alpha,\beta} - u_{\beta,\alpha} - \dot{u}_\alpha u_\beta + \dot{u}_\beta u_\alpha), \qquad w^\alpha = -(1/\sqrt{-g})\varepsilon^{\alpha\beta\gamma\delta} u_\beta \omega_{\gamma\delta}. \tag{1.10}$$

If $\omega_{\alpha\beta} \neq 0$ and $\dot{u}^\alpha = 0$ (what is assumed throughout), then necessarily the pressure $p =$ const and $\kappa p$ may be interpreted as the cosmological constant ($\kappa := 8\pi G/c^4$).

If any Killing vector field exists on a manifold (on which all the assumptions specified so far are fulfilled), then, in the coordinates of (1.9), it must be of the form:

$$k^\alpha = (C + \phi - y\phi,_y)\delta^\alpha{}_0 + \phi,_y \delta^\alpha{}_1 - \phi,_x \delta^\alpha{}_2 + \lambda\delta^\alpha{}_3, \tag{1.11}$$

where $C$ is an arbitrary constant and $\phi(x, y)$ and $\lambda(x, y)$ are arbitrary functions of two coordinates. Whenever $\phi,_\alpha \neq 0$, a transformation of the class (1.2) - (1.4) can be found that leads to:

$$k^\alpha = \delta^\alpha{}_1. \tag{1.12}$$

The metric then becomes independent of $x$, and the coordinates preserving (1.12) are determined up to the transformations:

$$t' = t - \int yH_{,y}\, dy + A, \qquad x' = x + H(y), \qquad y' = y, \qquad z' = z + T(y), \tag{1.13}$$

where $A$ is an arbitrary constant and $H, T$ are arbitrary functions.

The condition $\phi_{,\alpha} \neq 0$ that allows one to fulfil (1.12) means that the Killing vector $k^\alpha$ is linearly independent of the vectors $u^\alpha$ and $w^\alpha$ at every point of the spacetime region under consideration. In Paper 1, solutions of the Killing equations and of the Einstein equations were considered under the assumption that there exist three Killing vector fields on the manifold, two of which have $\phi =$ const in (1.11), while the third one has $\phi_{,\alpha} \neq 0$ and can be transformed to the form (1.12). In Paper 2, it was assumed that only one Killing field has $\phi =$ const, while two have $\phi_{,\alpha} \neq 0$. In the present paper, all three Killing fields will be assumed to have $\phi_{,\alpha} \neq 0$. One of them ($k_{(1)}$) can be transformed to the simple form (1.12), while the remaining two will have the general form (1.11).

In this Paper 3, no progress was made with the Einstein equations in any of the cases. Also, no related results were found in the literature except in case 1.1.2.2, see at the end of sec. V.

## II. The Lie algebra of the symmetry group.

According to the assumptions made in the preceding section, there exist the following three Killing vector fields:

$$\begin{aligned} k^\alpha_{(1)} &= \delta^\alpha_1, \\ k^\alpha_{(2)} &= (C_2 + \phi - y\phi_{,y})\delta^\alpha_0 + \phi_{,y}\, \delta^\alpha_1 - \phi_{,x}\, \delta^\alpha_2 + \lambda_2(x,y)\delta^\alpha_3, \\ k^\alpha_{(3)} &= (C_3 + \psi - y\psi_{,y})\delta^\alpha_0 + \psi_{,y}\, \delta^\alpha_1 - \psi_{,x}\, \delta^\alpha_2 + \lambda_3(x,y)\delta^\alpha_3, \end{aligned} \tag{2.1}$$

where $C_2$ and $C_3$ are arbitrary constants, and $\phi$, $\psi$, $\lambda_2$ and $\lambda_3$ are unknown functions of $(x, y)$, to be determined from the commutation relations. The coordinates of (2.1) are determined up to (1.13).

The fields $k_{(1)}, k_{(2)}$ and $k_{(3)}$ will form a Lie algebra if constants $a, \ldots, j$ exist such that:

$$\begin{aligned} [k_{(1)}, k_{(2)}] &= ak_{(1)} + bk_{(2)} + ck_{(3)}, \\ [k_{(1)}, k_{(3)}] &= dk_{(1)} + ek_{(2)} + fk_{(3)}, \\ [k_{(2)}, k_{(3)}] &= gk_{(1)} + hk_{(2)} + jk_{(3)}, \end{aligned} \tag{2.2}$$

Eqs. (2.2) are equivalent to the following set:

$$\phi_{,x} - y\phi_{,xy} = b(C_2 + \phi - y\phi_{,y}) + c(C_3 + \psi - y\psi_{,y}), \tag{2.3a}$$

$$\phi_{,xy} = a + b\phi_{,y} + c\psi_{,y}, \tag{2.3b}$$

$$\phi_{,xx} = b\phi_{,x} + c\psi_{,x}, \tag{2.3c}$$

$$\lambda_{2,x} = b\lambda_2 + c\lambda_3, \tag{2.3d}$$

$$\psi_{,x} - y\psi_{,xy} = e(C_2 + \phi - y\phi_{,y}) + f(C_3 + \psi - y\psi_{,y}), \tag{2.3e}$$

$$\psi_{,xy} = d + e\phi_{,y} + f\psi_{,y}\,, \tag{2.3f}$$

$$\psi_{,xx} = e\phi_{,x} + f\psi_{,x}\,, \tag{2.3g}$$

$$\lambda_{3,x} = e\lambda_2 + f\lambda_3, \tag{2.3h}$$

$$\phi_{,y}(\psi_{,x} - y\psi_{,xy}) + y\phi_{,x}\psi_{,yy} - \psi_{,y}(\phi_{,x} - y\phi_{,xy}) - y\psi_{,x}\phi_{,yy}$$

$$= h(C_2 + \phi - y\phi_{,y}) + j(C_3 + \psi - y\psi_{,y}), \tag{2.3i}$$

$$\phi_{,y}\psi_{,xy} - \phi_{,x}\psi_{,yy} - \psi_{,y}\phi_{,xy} + \psi_{,x}\phi_{,yy} = g + h\phi_{,y} + j\psi_{,y}\,, \tag{2.3j}$$

$$-\phi_{,y}\psi_{,xx} + \phi_{,x}\psi_{,xy} + \psi_{,y}\phi_{,xx} - \psi_{,x}\phi_{,xy} = -h\phi_{,x} - j\psi_{,x}\,, \tag{2.3k}$$

$$\phi_{,y}\lambda_{3,x} - \phi_{,x}\lambda_{3,y} - \psi_{,y}\lambda_{2,x} + \psi_{,x}\lambda_{2,y} = h\lambda_2 + j\lambda_3. \tag{2.3l}$$

The equations (2.3a-c) are integrated with the result:

$$\phi_{,x} = ay + b\phi + c\psi + bC_2 + cC_3; \tag{2.4a}$$

the equations (2.3e-g) are integrated with the result:

$$\psi_{,x} = dy + e\phi + f\psi + eC_2 + fC_3; \tag{2.4b}$$

and the equations (2.3i-k) are integrated with the result:

$$\phi_{,y}\psi_{,x} - \phi_{,x}\psi_{,y} = gy + h\phi + j\psi + hC_2 + jC_3. \tag{2.4c}$$

The equations are now sorted as follows. Eqs. (2.3d) and (2.3h) form a set that determines $\lambda_2$ and $\lambda_3$, eqs. (2.4a-b) form a set that determines $\phi$ and $\psi$, the remaining two equations ((2.4c) and (2.3l)) are consistency conditions to be imposed on the solutions of the two former sets.

Note that the set (2.4a-b) and the set {(2.3d), (2.3h)} are of the same general form: each of them is an (ordinary differential) linear vector equation of first order:

$$U_{,x} = AU + W, \tag{2.5}$$

where, for (2.4a-b), the constant matrix $A$ and the vectors $U$ and $W$ are:

$$A = \begin{pmatrix} b & c \\ e & f \end{pmatrix}, \qquad U = \begin{pmatrix} \phi \\ \psi \end{pmatrix}, \qquad W = y\begin{pmatrix} a \\ d \end{pmatrix} + A\begin{pmatrix} C_2 \\ C_3 \end{pmatrix}, \tag{2.6}$$

while the set {(2.3d), (2.3h)} is homogeneous, so $W = 0$ in (2.5), the matrix $A$ is the same as in (2.6) and $U = \begin{pmatrix} \lambda_2 \\ \lambda_3 \end{pmatrix}$.

With the constants $b$, $c$, $e$ and $f$ being all arbitrary, several cases will have to be considered separately. Just as in Paper 2, the cases that arise will be organized into a binary tree an numbered in a positional system that will enable one to quickly identify the complementary part of each alternative (see also the diagram).

The first alternative appears in solving the characteristic equation for the matrix $A$. Its eigenvalues are:

$$\alpha_{1,2} = \frac{1}{2}(b + f + \varepsilon_{1,2}\sqrt{\Delta}), \tag{2.7}$$

where:

$$\Delta := (b - f)^2 + 4ce, \qquad \varepsilon_1 = 1, \varepsilon_2 = -1. \tag{2.8}$$

We first consider:

**Case 1: $\Delta \neq 0$ (i.e. $A$ has two distinct eigenvalues).**

The second alternative appears immediately in finding the eigenvectors of $A$: the cases $c \neq 0$ and $c = 0$ have to be considered separately.

**Case 1.1:** $c \neq 0$.

The solution of the set $\{$(2.3d), (2.3h)$\}$ is then:

$$\begin{aligned} \lambda_2 &= 2cL_2(y)e^{\alpha_1 x} + 2cL_3(y)e^{\alpha_2 x}, \\ \lambda_3 &= (f - b + \sqrt{\Delta})L_2(y)e^{\alpha_1 x} + (f - b - \sqrt{\Delta})L_3(y)e^{\alpha_2 x}, \end{aligned} \tag{2.9}$$

where $L_2(y)$ and $L_3(y)$ are arbitrary functions. The cases $\Delta > 0$ and $\Delta < 0$ could be considered together for a large part of the reasoning. When $\Delta < 0, \alpha_1$ and $\alpha_2$ are complex and $\alpha_2 = \overline{\alpha_1}$. Then, $L_2$ and $L_3$ have to be complex, too, with $L_3 = \overline{L_2}$. However, the two cases lead to different sets of Bianchi types, and so it will be convenient to split them here.

**Case 1.1.1: $\Delta > 0$ (i.e. both eigenvalues are real).**

Then, in solving the set (2.4a-b), the cases $\det A \neq 0$ and $\det A = 0$ have to be considered separately. In the end, however, the case $\det A = 0$ turns out to be empty, i.e. in all subcases that arise in it there exists a linear combination of the Killing vectors $k_{(2)}$ and $k_{(3)}$ with constant coefficients that is spanned on $u$ and $w$. This means that all these subcases are in the domain of Paper 2 and need not be considered here. Therefore we will do away with the case $\det A = 0$ by only indicating the method of verification of the statement above.

When $\det A = 0$, the following is true (from (2.7) - (2.8)):

$$e = bf/c, \qquad \Delta = (b + f)^2, \qquad \alpha_1 = b + f := \alpha, \qquad \alpha_2 = 0. \tag{2.10}$$

Since we are still in Case 1 in which $\alpha_2 \neq \alpha_1$ by assumption, we can take it for granted that $\alpha \neq 0$ here. The solutions of (2.3d), (2.3h) and (2.4a-b) are:

$$\begin{aligned} \lambda_2 = 2c[L_2(y) + L_3(y)e^{\alpha x}], &\qquad \lambda_3 = 2fL_2(y) - 2bL_3(y)e^{\alpha x}, \\ \phi = F(y)e^{\alpha x} + (af - cd)&xy/\alpha + P(y) - (ab + cd)y/\alpha^2 - C_2, \\ \psi = (f/c)F(y)e^{\alpha x} - (b/c)&(af - cd)xy/\alpha - (b/c)P(y) \\ &-(f/c)(ab + cd)y/\alpha^2 - C_3, \end{aligned} \tag{2.11}$$

where $F(y)$ and $P(y)$ are other arbitrary functions. The further procedure goes exactly as for the case $\det A \neq 0$ presented below, and leads to the results specified above.

From now on, in Case 1.1.1 we assume that:

$$\det A = bf - ce \neq 0. \tag{2.12}$$

Then the solution of (2.4a-b) is:

$$\phi = 2cF(y)e^{\alpha_1 x} + 2cP(y)e^{\alpha_2 x} - \frac{af - cd}{bf - ce}y - C_2,$$

$$\psi = (f - b + \sqrt{\Delta})F(y)e^{\alpha_1 x} + (f - b - \sqrt{\Delta})P(y)e^{\alpha_2 x} - \frac{-ae + bd}{bf - ce}y - C_3. \tag{2.13}$$

The solutions (2.9) and (2.13) must now obey the consistency conditions (2.4$c$) and (2.3$l$). Eq. (2.4$c$) is a polynomial in $e^{\alpha_1 x}$ and $e^{\alpha_2 x}$ whose coefficients are functions of $y$, the polynomial contains $e^{(\alpha_1+\alpha_2)x}, e^{\alpha_1 x}, e^{\alpha_2 x}$ and terms independent of $x$. The cases $\alpha_2 \neq -\alpha_1$ and $\alpha_2 = -\alpha_1$ require separate consideration. We first consider:

**Case 1.1.1.1:** $\alpha_2 \neq -\alpha_1$.

Comparison of coefficients of $e^{(\alpha_1+\alpha_2)x}$ on both sides of (2.4$c$) leads to:

$$-\alpha_2 F_{,y} P + \alpha_1 F P_{,y} = 0. \tag{2.14}$$

If $F = 0$, then (2.14) is fulfilled identically, and this case has to be considered separately. The result is similar as in the case $\det A = 0$. The consideration, parallel to the one that follows below, reveals that with $F = 0$ either one of the Killing fields becomes collinear with rotation (and this situation is in the domain of Paper 2) or the symmetry group becomes two dimensional (which case is not considered here at all). Consequently, we proceed assuming $F \neq 0$. The solution of (2.14) is then:

$$P = \beta F^{\alpha_2/\alpha_1}, \tag{2.15}$$

where $\beta$ is an arbitrary constant; $\alpha_1 \neq 0$ because $\det A \neq 0$.

The coefficients of $Fe^{\alpha_1 x}$ on both sides of (2.4$c$) imply:

$$h = [\alpha_1/(bf - ce)][-\frac{1}{2c}(af - cd)(f - b + \sqrt{\Delta}) - ae + bd] - \frac{j}{2c}(f - b + \sqrt{\Delta}). \tag{2.16}$$

In considering the coefficients of $e^{\alpha_2 x}$ in (2.4$c$) we have to set aside the case $P = 0$ for separate consideration because the terms with $e^{\alpha_2 x}$ all vanish identically when $P = 0$. However, the case $P = 0$ is in fact empty in the same sense as the case $F = 0$: either the Killing field $k_{(3)}$ becomes collinear with rotation, and this situation is in the domain of Paper 2, or the symmetry group becomes two-dimensional. Hence, we shall follow the case $P \neq 0$ only. Then the coefficients of $Pe^{\alpha_2 x}$ on both sides of (2.4$c$) imply:

$$j = (bf - ce)^{-1}[-(af - cd)f + (-ae + bd)c], \tag{2.17}$$

and the terms independent of $x$ imply:

$$g = (bf - ce)^{-1}[-(af - cd)d + (-ae + bd)a]. \tag{2.18}$$

Eq. (2.3$l$) is a polynomial in exponential functions of $x$ that involves $e^{2\alpha_1 x}$, $e^{2\alpha_2 x}$, $e^{(\alpha_1+\alpha_2)x}$, $e^{\alpha_1 x}$ and $e^{\alpha_2 x}$. In consequence of (2.9) - (2.18) most of the resulting equations are fulfilled identically, the only one that brings in new information is:

$$-2\alpha_2 F_{,y} L_3 + 2\alpha_1 F L_{3,y} + 2(\beta\alpha_2 F^{\alpha_2/\alpha_1})(L_2 F_{,y}/F - L_{2,y}) = 0. \tag{2.19}$$

This is integrated with the result:

$$L_3 = (1/\alpha_1) F^{\alpha_2/\alpha_1}(\gamma + \beta\alpha_2 L_2/F), \tag{2.20}$$

where $\gamma$ is a new arbitrary constant.

Finally, the following Killing fields resulted:

$$k^\alpha_{(1)} = \delta^\alpha_1,$$

$$k^\alpha_{(2)} = [2c(F - yF_{,y})e^{\alpha_1 x} + 2c\beta F^{\alpha_2/\alpha_1}(1 - \frac{\alpha_2}{\alpha_1} yF_{,y}/F)e^{\alpha_2 x}]\delta^\alpha_0$$
$$+(2cF_{,y} e^{\alpha_1 x} + 2c\beta\frac{\alpha_2}{\alpha_1} F^{\alpha_2/\alpha_1 - 1} F_{,y} e^{\alpha_2 x} - \frac{af - cd}{bf - ce})\delta^\alpha_1$$
$$-(2c\alpha_1 F e^{\alpha_1 x} + 2c\alpha_2\beta F^{\alpha_2/\alpha_1} e^{\alpha_2 x})\delta^\alpha_2 + [2cL_2 e^{\alpha_1 x} + 2\frac{c}{\alpha_1} F^{\alpha_2/\alpha_1}(\gamma + \beta\alpha_2 L_2/F)e^{\alpha_2 x}]\delta^\alpha_3,$$

$$k^\alpha_{(3)} = [(f - b + \sqrt{\Delta})(F - yF_{,y})e^{\alpha_1 x} + (f - b - \sqrt{\Delta})\beta F^{\alpha_2/\alpha_1}(1 - \frac{\alpha_2}{\alpha_1} yF_{,y}/F)e^{\alpha_2 x}]\delta^\alpha_0$$
$$+[(f - b + \sqrt{\Delta})F_{,y} e^{\alpha_1 x} + (f - b - \sqrt{\Delta})\beta\frac{\alpha_2}{\alpha_1} F^{\alpha_2/\alpha_1 - 1} F_{,y} e^{\alpha_2 x} - \frac{-ae + bd}{bf - ce}]\delta^\alpha_1$$
$$-[(f - b + \sqrt{\Delta})\alpha_1 F e^{\alpha_1 x} + (f - b - \sqrt{\Delta})\beta\alpha_2 F^{\alpha_2/\alpha_1} e^{\alpha_2 x}]\delta^\alpha_2$$
$$+[(f - b + \sqrt{\Delta})L_2 e^{\alpha_1 x} + (f - b - \sqrt{\Delta})\frac{1}{\alpha_1} F^{\alpha_2/\alpha_1}(\gamma + \beta\alpha_2 L_2/F)e^{\alpha_2 x}]\delta^\alpha_3. \tag{2.21}$$

These formulae are simplified by changing the basis in the algebra of the Killing vectors. Taking $k'^\alpha_{(2)} = k^\alpha_{(2)} + \frac{af - cd}{bf - ce} k^\alpha_{(1)}$ and $k'^\alpha_{(3)} = k^\alpha_{(3)} + \frac{-ae + bd}{bf - ce} k^\alpha_{(1)}$ instead of $k^\alpha_{(2)}$ and $k^\alpha_{(3)}$ we obtain the same result as if:

$$a = d = 0. \tag{2.22}$$

With (2.22), we take $k'^\alpha_{(3)} = (2\sqrt{\Delta})^{-1}[k^\alpha_{(3)} - \frac{1}{2c}(f - b - \sqrt{\Delta})k^\alpha_{(2)}]$ instead of $k^\alpha_{(3)}$ and $k'^\alpha_{(2)} = (2c\beta)^{-1}(k^\alpha_{(2)} - 2ck'^\alpha_{(3)})$ instead of $k^\alpha_{(2)}$. Further simplification results from the transformation (1.13) with:

$$H = \alpha_1^{-1} \ln F, \qquad T_{,y} = L_2/(\alpha_1 F). \tag{2.23}$$

The Killing field $k^\alpha_{(1)}$ does not change, while the other two become (all primes dropped):

$$k^\alpha_{(2)} = e^{\alpha_2 x}\{\delta^\alpha_0 - \alpha_2\delta^\alpha_2 + [\gamma/(\beta\alpha_1)]\delta^\alpha_3\}, \qquad k^\alpha_{(3)} = e^{\alpha_1 x}(\delta^\alpha_0 - \alpha_1\delta^\alpha_2). \tag{2.24}$$

Using (2.22) in (2.16) - (2.18) we obtain:

$$g = h = j = 0. \tag{2.25}$$

Looking at the commutation relations one can see that in effect we have also achieved $c = e = 0$, $\alpha_1 = b$, $\alpha_2 = f$, even though the initial basis (2.21) was calculated under the assumption $c \neq 0$. This allows us to predict that in the case $c = 0$, set aside for separate consideration, we will obtain (2.24) again; see case 1.2 in sec. VI. In view of the assumptions made earlier, we have:

$$b + f \neq 0 \neq b - f, \qquad b \neq 0 \neq f, \tag{2.26}$$

and consequently the Bianchi type is $\mathrm{VI}_h$, with the free parameter being $a_B = (b+f)/(b-f)$.

The Killing fields are simplified even further after the following transformation that leads out of the Plebański class:

$$t' = (b-f)^{-1}e^{-fx}(bt+y), \qquad x' = x, \qquad y' = (b-f)^{-1}be^{-bx}(ft+y),$$
$$z' = -\gamma[\beta(b-f)]^{-1}(t + y/b) + z, \tag{2.27}$$

that results in (primes dropped):

$$k_{(1)}^{\alpha} = -ft\delta_0^{\alpha} + \delta_1^{\alpha} - by\delta_2^{\alpha}, \qquad k_{(2)}^{\alpha} = \delta_0^{\alpha}, \qquad k_{(3)}^{\alpha} = \delta_2^{\alpha},$$
$$u^{\alpha} = (b-f)^{-1}[be^{-fx}\delta_0^{\alpha} + bfe^{-bx}\delta_2^{\alpha} - (\gamma/\beta)\delta_3^{\alpha}], \qquad w^{\alpha} = n\delta_3^{\alpha}. \tag{2.28}$$

In the new coordinates, the Killing equations imply for the metric:

$$g_{00} = e^{2fx}[1 - (f/b)^2 + \gamma^2 g_{33}/(b\beta)^2 - 2f\gamma h_{23}/(b\beta) + f^2 h_{22}],$$
$$g_{01} = e^{fx}[\gamma g_{13}/(b\beta) - fh_{12}], \qquad g_{02} = e^{(b+f)x}[\gamma h_{23}/(b\beta) - (b-f)/b^2 - fh_{22}],$$
$$g_{03} = e^{fx}[\gamma g_{33}/(b\beta) - fh_{23}], \qquad g_{12} = e^{bx}h_{12}, \qquad g_{22} = e^{2bx}h_{22}, \qquad g_{23} = e^{bx}h_{23}, \tag{2.29}$$

where $g_{11}$, $h_{12}$, $g_{13}$, $h_{22}$, $h_{23}$ and $g_{33}$ are arbitrary functions of $z$.

## III. Case 1.1.1.2: $\alpha_2 = -\alpha_1$.

All equations up to (2.13) still apply, but (2.4$c$) has to be reconsidered. With $\alpha_2 = -\alpha_1$ the following equations hold:

$$f = -b, \qquad \Delta = 4(b^2 + ce), \qquad \alpha_1 = \sqrt{\Delta}/2 = -\alpha_2. \tag{3.1}$$

In considering eqs. (2.4$c$) and (2.3$l$) it may be assumed that $F \neq 0 \neq P$ because the opposite cases lead, just as before, out of the domain of this paper. With $F \neq 0 \neq P$, the coefficients of $e^{\alpha_1 x}$, of $e^{-\alpha_1 x}$ and the terms independent of $x$ (which now include the coefficients of $e^{(\alpha_1+\alpha_2)x}$) in (2.4$c$) imply, respectively:

$$h = \Delta^{-1/2}[-(ab/c + d)(-2b + \sqrt{\Delta}) - 2(-ae + bd)] - j(-2b + \sqrt{\Delta})/(2c),$$
$$j = a, \qquad c\Delta FP = (1/2)\{g + \Delta^{-1}[-h(ab + cd) + a(-ae + bd)]\}y^2 + B, \tag{3.2}$$

where $B$ is an arbitrary constant.

In eq. (2.3$l$) only the terms independent of $x$ provide a piece of new information, the other parts of the equation are fulfilled identically in consequence of (3.1) - (3.2). The new information is:

$$L_3 = \gamma/(\alpha_1 F) - PL_2/F \tag{3.3}$$

(the integration constant $\gamma$ was chosen so as to correspond to (2.20)). Like in sec. II, we first construct the Killing fields by substituting (3.1) - (3.3), (2.9) and (2.13) into (2.1), then simplify the result by changing the basis in the Lie algebra and carrying out the coordinate transformations (1.13) (the difference is that here the coefficient in the formula for $k'^{\alpha}_{(2)}$ is $\Delta$ instead of $(2c\beta)^{-1}$). The result is:

$$a = d = 0, \qquad k^{\alpha}_{(1)} = \delta^{\alpha}_1, \qquad k^{\alpha}_{(3)} = e^{\alpha_1 x}(\delta^{\alpha}_0 - \alpha_1 \delta^{\alpha}_2),$$
$$k^{\alpha}_{(2)} = e^{-\alpha_1 x}[(-gy^2 + 2B)\delta^{\alpha}_0 + 2gy\delta^{\alpha}_1 + \alpha_1(gy^2 + 2B)\delta^{\alpha}_2 + 2(c\gamma\Delta/\alpha_1)\delta^{\alpha}_3]. \tag{3.4}$$

The commutation relations are:

$$[k_{(1)}, k_{(2)}] = -\alpha_1 k_{(2)}, \qquad [k_{(2)}, k_{(3)}] = 2g\alpha_1 k_{(1)}, \qquad [k_{(3)}, k_{(1)}] = -\alpha_1 k_{(3)}, \tag{3.5}$$

and they correspond to the Bianchi type VIII when $g \neq 0$ and type $\text{VI}_0$ when $g = 0$ (note that $\alpha_1 \neq 0$ by the assumption defining case 1.1.1).

The Killing fields are further simplified by the coordinate transformation:

$$t' = t + y/\alpha_1 - [2B\alpha_1/(c\gamma\Delta)]z, \qquad x' = x, \qquad y' = y, \qquad z' = [\alpha_1/(2c\gamma\Delta)]z \tag{3.6}$$

that results (with primes dropped) in:

$$k^{\alpha}_{(1)} = \delta^{\alpha}_1, \qquad k^{\alpha}_{(2)} = e^{-\alpha_1 x}[2gy\delta^{\alpha}_1 + \alpha_1(gy^2 + 2B)\delta^{\alpha}_2 + \delta^{\alpha}_3], \qquad k^{\alpha}_{(3)} = -\alpha_1 e^{\alpha_1 x}\delta^{\alpha}_2,$$

$$u^{\alpha} = \delta^{\alpha}_0, \qquad w^{\alpha} = [n\alpha_1/(2c\gamma\Delta)](-4B\delta^{\alpha}_0 + \delta^{\alpha}_3),$$

$$g_{00} = 1, \qquad g_{01} = y, \qquad g_{02} = -1/\alpha_1, \qquad g_{03} = 4B, \tag{3.7}$$

the other components of $g_{ij}$ being just unknown functions. The Killing equations for $k^{\alpha}_{(1)}$ now imply that the metric is independent of $x$, while those for $k^{\alpha}_{(3)}$ are solved by:

$$g_{11} = (\alpha_1 y)^2 g_{22} - 2\alpha_1 y H_{12} + H_{11},$$
$$g_{12} = -\alpha_1 y g_{22} + H_{12}, \qquad g_{13} = -\alpha_1 y g_{23} + H_{13}, \tag{3.8}$$

where $H_{11}$, $H_{12}$, $H_{13}$, $g_{22}$, $g_{23}$ and $g_{33}$ are arbitrary functions of $t$ and $z$.

In solving the Killing equations for $k^{\alpha}_{(2)}$, three cases have to be considered separately:

**Case I:** $gB \neq 0$.

In fact, the cases $gB > 0$ and $gB < 0$ lead to different results, but the formulae for $gB < 0$ can be easily reconstructed from those for $gB > 0$ by taking real combinations of the complex solutions. Hence, we shall only present the formulae for $gB > 0$. The solution of the Killing equations is:

$$\mu := (2gB)^{1/2}\alpha_1, \qquad U := h_{12}\cos(4\mu z) - k_{12}\sin(4\mu z),$$
$$V := h_{13}\cos(2\mu z) - k_{13}\sin(2\mu z)$$
$$H_{11} = -(2B\alpha_1^2/\mu)U + h_{11}, \qquad H_{12} = h_{12}\sin(4\mu z) + k_{12}\cos(4\mu z) - H_{13}/(4B\alpha_1),$$
$$H_{13} = h_{13}\sin(2\mu z) + k_{13}\cos(2\mu z), \qquad g_{22} = (g/\mu)U - [g/(2B\alpha_1\mu)]V + h_{22},$$
$$g_{23} = [\mu/(2B\alpha_1^2)]V - h_{33}/(4B\alpha_1), \qquad g_{33} = h_{33}, \tag{3.9}$$

where the $h_{ij}(t)$ and $k_{ij}(t)$ are arbitrary functions.

When $gB < 0$, $\mu$ is imaginary. Then the trigonometric functions go over into the appropriate hyperbolic functions, and $h_{12}$ and $h_{13}$ have to be taken imaginary, too.

**Case II:** $B = 0$.

The solution of the Killing equations for $k_{(2)}^{\alpha}$ is then:

$$H_{11} = h_{33}(\alpha_1 z)^2 + 2\alpha_1 h_{13} z + h_{11},$$
$$H_{12} = -\alpha_1^2 g h_{33} z^3 - 3\alpha_1 g h_{13} z^2 + (\alpha_1 h_{23} - 2gh_{11})z + h_{12}, \qquad H_{13} = \alpha_1 h_{33} z + h_{13},$$
$$g_{22} = (\alpha_1 g)^2 h_{33} z^4 + 4\alpha_1 g^2 h_{13} z^3 - 2g(\alpha_1 h_{23} - 2gh_{11})z^2 - 4gh_{12}z + h_{22},$$
$$g_{23} = -\alpha_1 g h_{33} z^2 - 2gh_{13}z + h_{23}, \qquad g_{33} = h_{33}, \tag{3.10}$$

where the $h_{ij}(t)$ are arbitrary functions.

**Case III:** $g = 0$.

The Killing equations for $k_{(2)}^{\alpha}$ imply here:

$$H_{11} = [(4B\alpha_1)^2 h_{22} + 8B\alpha_1 h_{23} + h_{33}](\alpha_1 z)^2 + (8B\alpha_1 h_{12} + 2h_{13})\alpha_1 z + h_{11},$$
$$H_{12} = (4B\alpha_1 h_{22} + h_{23})\alpha_1 z + h_{12}, \qquad H_{13} = (4B\alpha_1 h_{23} + h_{33})\alpha_1 z + h_{13},$$
$$g_{22} = h_{22}, \qquad g_{23} = h_{23}, \qquad g_{33} = h_{33}, \tag{3.11}$$

where the $h_{ij}(t)$ are arbitrary functions.

With this, case 1.1.1 is exhausted and we go back to (2.9) to consider the other branch of the alternative.

## IV. Case 1.1.2: $\Delta < 0$ (i.e. the eigenvalues $\alpha_1$ and $\alpha_2$ are complex and conjugate to each other).

As stated before, it is more convenient to reparametrize (2.9) so that it contains only real quantities, and then repeat the procedure of case 1.1.1 in this new parametrization. We denote:

$$\sqrt{-\Delta} = D, \qquad F = G + iJ, \qquad P = G - iJ, \qquad L_2 = M + iN, \qquad L_3 = M - iN, \tag{4.1}$$

where $G, J, M$ and $N$ are new unknown functions of $y$, and then:

$$\alpha_{1,2} = \frac{1}{2}(b+f) + \frac{1}{2}i\varepsilon_{1,2}D, \qquad \varepsilon_1 = 1, \qquad \varepsilon_2 = -1. \tag{4.2}$$

In this notation, eqs. (2.9) and (2.13) adapted to the case $\Delta < 0$ are:

$$\lambda_2 = 4ce^{(b+f)x/2}[M\cos(Dx/2) - N\sin(Dx/2)],$$
$$\lambda_3 = \frac{1}{2c}(f-b)\lambda_2 - 2D\mathrm{e}^{(b+f)x/2}[M\sin(Dx/2) + N\cos(Dx/2)],$$

$$\phi = 4ce^{(b+f)x/2}[G\cos(Dx/2) - J\sin(Dx/2)] - \frac{af-cd}{bf-ce}y - C_2,$$
$$\psi = 2(f-b)e^{(b+f)x/2}[G\cos(Dx/2) - J\sin(Dx/2)]$$
$$-2D\mathrm{e}^{(b+f)x/2}[G\sin(Dx/2) + J\cos(Dx/2)] - \frac{-ae+bd}{bf-ce}y - C_3. \tag{4.3}$$

Just as in Case 1.1.1, these expressions must now satisfy the consistency conditions (2.4$c$) and (2.3$l$). However, in considering them, the cases $b+f \neq 0$ and $b+f=0$ have to be taken separately.

**Case 1.1.2.1:** $b+f \neq 0$.

Both sides of (2.4$c$) are then polynomials in $e^{(b+f)x/2}$, and some of their coefficients involve $\cos(Dx/2)$ and $\sin(Dx/2)$. The coefficient of $e^{(b+f)x}$ leads to the equation:

$$4cD[-D(GG,_y + JJ,_y) + (b+f)(GJ,_y - G,_y J)] = 0. \tag{4.4}$$

We are working in the case $c \neq 0 \neq D$, so only the expression in square brackets can vanish. Its form suggests the substitution:

$$G = K\cos L, \qquad J = K\sin L, \tag{4.5}$$

where $K$ and $L$ are new functions of $y$. Eq. (4.4) becomes then $-DKK,_y + (b+f)K^2L,_y = 0$, and its solution is:

$$K = Be^{(b+f)L/D}, \tag{4.6}$$

where $B$ is an arbitrary constant; $B \neq 0$ or else we are back in the domain of Paper 2. Eqs. (4.5) and (4.6) provide a parametric representation of $G$ and $J$ in terms of the function $L$, which is arbitrary at this stage.

The coefficients of $e^{(b+f)x/2}$ in (2.4$c$) involve sin and cos that always go in fixed pairs. The coefficients of $\{e^{(b+f)x/2}[G\cos(Dx/2) - J\sin(Dx/2)]\}$ imply:

$$h = [4c(bf-ce)]^{-1}[(af-cd)(b^2-f^2+D^2) + 2c(b+f)(-ae+bd)] - j(f-b)/(2c), \tag{4.7}$$

and the coefficients of $\{e^{(b+f)x/2}[G\sin(Dx/2) + J\cos(Dx/2)]\}$ imply:

$$j = -(bf-ce)^{-1}[f(af-cd) - c(-ae+bd)]. \tag{4.8}$$

Finally, the terms independent of $x$ imply:

$$g = -(bf - ce)^{-1}[h(af - cd) + j(-ae + bd)]. \tag{4.9}$$

With (4.7) - (4.8), eq. (2.3$l$) is reduced to:

$$\begin{aligned} &4cD\mathrm{e}^{(b+f)x}[-D(GM,_y + G,_y M + JN,_y + J,_y N) \\ &+(b+f)(GN,_y - G,_y N - JM,_y + J,_y M)] = 0. \end{aligned} \tag{4.10}$$

Only the expression in square brackets can vanish in the case now considered. After (4.5) and (4.6) are substituted into (4.10), the resulting equation integrates to:

$$e^{-(b+f)L/D}\{-M[D\cos L + (b+f)\sin L] + N[(b+f)\cos L - D\sin L]\} = \gamma B(b+f) = \text{ const.} \tag{4.11}$$

Since $b + f \neq 0 \neq D$ by assumption, this can be solved for $N$:

$$N = [(b+f)\cos L - D\sin L]^{-1}\{\gamma B(b+f)e^{-(b+f)L/D} + M[D\cos L + (b+f)\sin L]\}. \tag{4.12}$$

The resulting Killing fields are:

$$k^{\alpha}_{(1)} = \delta^{\alpha}_{1},$$

$$\begin{aligned} k^{\alpha}_{(2)} &= 4ce^{(b+f)x/2}[(G - yG,_y)\cos(Dx/2) - (J - yJ,_y)\sin(Dx/2)]\delta^{\alpha}_{0} \\ &+\{4ce^{(b+f)x/2}[G,_y\cos(Dx/2) - J,_y\sin(Dx/2)] - \frac{af - cd}{bf - ce}\}\delta^{\alpha}_{1} \\ &-2ce^{(b+f)x/2}\{(b+f)[G\cos(Dx/2) - J\sin(Dx/2)] - D[G\sin(Dx/2) + J\cos(Dx/2)]\}\delta^{\alpha}_{2} \\ &+4ce^{(b+f)x/2}[M\cos(Dx/2) - N\sin(Dx/2)]\delta^{\alpha}_{3}, \end{aligned}$$

$$\begin{aligned} k^{\alpha}_{(3)} &= 2e^{(b+f)x/2}\{(f-b)[(G - yG,_y)\cos(Dx/2) - (J - yJ,_y)\sin(Dx/2)] \\ &-D[(G - yG,_y)\sin(Dx/2) + (J - yJ,_y)\cos(Dx/2)]\}\delta^{\alpha}_{0} \\ &+2e^{(b+f)x/2}\{(f-b)[G,_y\cos(Dx/2) - J,_y\sin(Dx/2)] \\ &-D[G,_y\sin(Dx/2) + J,_y\cos(Dx/2)]\} - \frac{-ae + bd}{bf - ce}\}\delta^{\alpha}_{1} \\ &-e^{(b+f)x/2}\{(f^2 - b^2 - D^2)[G\cos(Dx/2) - J\sin(Dx/2)] \\ &-2fD[G\sin(Dx/2) + J\cos(Dx/2)]\}\delta^{\alpha}_{2} \\ &+e^{(b+f)x/2}\{2(f-b)[M\cos(Dx/2) - N\sin(Dx/2)] \\ &-2D[M\sin(Dx/2) + N\cos(Dx/2)]\}\delta^{\alpha}_{3}. \end{aligned} \tag{4.13}$$

Again, the formulae simplify when the basis in the Lie algebra is changed. First, we take $k'^{\alpha}_{(2)} = k^{\alpha}_{(2)} + \frac{af-cd}{bf-ce} k^{\alpha}_{(1)}$ and $k'^{\alpha}_{(3)} = k^{\alpha}_{(3)} + \frac{-ae+bd}{bf-ce} k^{\alpha}_{(1)}$ instead of $k^{\alpha}_{(2)}$ and $k^{\alpha}_{(3)}$ respectively. The result is equivalent to:

$$a = d = 0, \tag{4.14}$$

and then (4.7) - (4.9) simplify to:

$$g = h = j = 0. \tag{4.15}$$

With (4.14) taken into account, we change the basis again by taking $k'^{\alpha}_{(3)} = (-2BD)^{-1}(k^{\alpha}_{(3)} - \frac{f-b}{2c} k^{\alpha}_{(2)})$ and $k'^{\alpha}_{(2)} = (4cB)^{-1} k^{\alpha}_{(2)}$ instead of $k^{\alpha}_{(3)}$ and $k^{\alpha}_{(2)}$, and carry out the transformation (1.13) with:

$$H = 2L/D, \qquad T_{,y} = 2[B(b+f)]^{-1} e^{-(b+f)L/D} (M \cos L + N \sin L) \tag{4.16}$$

The result is equivalent to $L = 0 = M$, i.e.:

$$G = B, \qquad J = 0, \qquad N = \gamma B. \tag{4.17}$$

The resulting Killing fields are:

$$\begin{aligned} k^{\alpha}_{(1)} &= \delta^{\alpha}_{1}, \\ k^{\alpha}_{(2)} &= e^{(b+f)x/2} [\cos(Dx/2) \delta^{\alpha}_{0} - \frac{1}{2} W \delta^{\alpha}_{2} - \gamma \sin(Dx/2) \delta^{\alpha}_{3}], \\ k^{\alpha}_{(3)} &= e^{(b+f)x/2} [\sin(Dx/2) \delta^{\alpha}_{0} - \frac{1}{2} V \delta^{\alpha}_{2} + \gamma \cos(Dx/2) \delta^{\alpha}_{3}, \end{aligned} \tag{4.18}$$

where:

$$\begin{aligned} W &:= (b+f) \cos(Dx/2) - D \sin(Dx/2), \\ V &:= D \cos(Dx/2) + (b+f) \sin(Dx/2). \end{aligned} \tag{4.19}$$

The commutation relations are:

$$[k_1, k_2] = \frac{1}{2}(b+f) k_2 - \frac{1}{2} D k_3, \qquad [k_2, k_3] = 0, \qquad [k_3, k_1] = -\frac{1}{2} D k_2 - \frac{1}{2}(b+f) k_3, \tag{4.20}$$

and they correspond to Bianchi type $\mathrm{VII}_h$ with the free parameter $a_B = -(b+f)/D$. Since $[k_2, k_3] = 0$, coordinates can be adapted to $k_2$ and $k_3$ simultaneously. The following transformation does it:

$$\begin{aligned} t' &= -D^{-1} e^{-(b+f)x/2} [Wt + 2\cos(Dx/2) y], \qquad x' = x, \\ y' &= e^{-(b+f)x/2} \{D^{-1} V t - 2W^{-1} [1 - D^{-1} \cos(Dx/2) V] y\}, \\ z' &= (\gamma/D)(b+f) t + 2(\gamma/D) y + z, \end{aligned} \tag{4.21}$$

but the new coordinates are no longer in the Plebański class. After the transformation, with primes dropped:

$$k^\alpha_{(1)} = \frac{1}{2}[-(b+f)t + Dy]\delta^\alpha_0 + \delta^\alpha_1 - \frac{1}{2}[Dt + (b+f)y]\delta^\alpha_2, \qquad k^\alpha_{(2)} = \delta^\alpha_2, \qquad k^\alpha_{(3)} = \delta^\alpha_0,$$

$$u^\alpha = D^{-1}e^{-(b+f)x/2}(-W\delta^\alpha_0 + V\delta^\alpha_2) + (\gamma/D)(b+f)\delta^\alpha_3, \qquad w^\alpha = n\delta^\alpha_3. \tag{4.22}$$

The formulae for $g_{0\alpha}$ ($\alpha = 0, 1, 2, 3$) in terms of $g_{ij}$ ($i, j = 1, 2, 3$) in the new coordinates are given in Appendix A.

In the new coordinates, the metric is independent of $t$ and $y$, while the Killing equations for $k^\alpha_{(1)}$ have the following solution:

$$g_{11} = h_{11}, \qquad g_{12} = e^{(b+f)x/2}[Wh_{12} - (\gamma/D)(b+f)\cos(Dx/2)h_{13}], \qquad g_{13} = h_{13},$$

$$g_{22} = e^{(b+f)x}\{[\gamma^2(b+f)^2h_{33}/D^2 + 1]\cos^2(Dx/2)$$

$$-2(\gamma/D)(b+f)\cos(Dx/2)Wh_{23} + W^2h_{22}\},$$

$$g_{23} = e^{(b+f)x/2}[Wh_{23} - (\gamma/D)(b+f)\cos(Dx/2)h_{33}], \qquad g_{33} = h_{33}, \tag{4.23}$$

where the $h_{ij}(z)$ are arbitrary functions. The components $g_{0\alpha}$ in terms of those given above are given in Appendix A.

**V. Case 1.1.2.2:** $f = -b$.

We go back to eqs. (4.3) which simplify as follows:

$$\lambda_2 = 4c[M\cos(Dx/2) - N\sin(Dx/2)],$$
$$\lambda_3 = -(b/c)\lambda_2 - 2D[M\sin(Dx/2) + N\cos(Dx/2)], \tag{5.1}$$

$$\phi = 4c[G\cos(Dx/2) - J\sin(Dx/2)] - \frac{ab+cd}{b^2+ce}y - C_2,$$
$$\psi = -4b[G\cos(Dx/2) - J\sin(Dx/2)]$$
$$-2D[G\sin(Dx/2) + J\cos(Dx/2)] + \frac{-ae+bd}{b^2+ce}y - C_3, \tag{5.2}$$

With these new forms of $\phi, \psi, \lambda_2$ and $\lambda_3$ we reconsider ($2.4c$) and ($2.3l$). Eq. ($2.4c$) is now linear inhomogeneous in $\sin(Dx/2)$ and $\cos(Dx/2)$. The coefficients of $\cos(Dx/2)$, of $\sin(Dx/2)$ and the terms independent of $x$ imply, respectively:

$$\frac{ab+cd}{b^2+ce}D^2G - 2aDJ = 4chG - 4bjG - 2jDJ, \tag{5.3}$$

$$-\frac{ab+cd}{b^2+ce}D^2J - 2aDG = -4chJ + 4bjJ - 2jDG, \tag{5.4}$$

$$-4cD^2(GG,_y + JJ,_y) = gy + (b^2 + ce)^{-1}[-h(ab + cd) + j(-ae + bd)]y. \qquad (5.5)$$

From (5.3) and (5.4) it follows that $2(j - a)D(G^2 + J^2) = 0$. Since $D \neq 0$ by assumption, and $G = J = 0$ leads to the domain of Paper 2, this implies:

$$j = a. \qquad (5.6)$$

With (5.6), eqs. (5.3) and (5.4) reduce to:

$$h = [4c(b^2 + ce)]^{-1}(ab + cd)D^2 + ab/c. \qquad (5.7)$$

The integral of (5.5) is:

$$-2cD^2(G^2 + J^2) = \{g + (b^2 + ce)^{-1}[-h(ab + cd) + a(-ae + bd)\}y^2/2 + B, \qquad (5.8)$$

where $B =$ const.

Eq. (2.3$l$) leads now to two additional equations, one of which has the solution:

$$GM + JN = \gamma/D = \text{ const}, \qquad (5.9)$$

and what remains of (2.3$l$) is then:

$$4aD[M\sin(Dx/2) + N\cos(Dx/2)] = 0. \qquad (5.10)$$

This has two solutions, $a = 0$ and $M = N = 0$, that must be considered separately. However, the case $M = N = 0$ turns out to be included as the subcase $\gamma = 0$ of the formulae below. Hence:

$$a = j = 0, \qquad h = \frac{1}{4}(b^2 + ce)^{-1}dD^2. \qquad (5.11)$$

The resulting Killing fields are:

$$k^\alpha_{(1)} = \delta^\alpha_1,$$

$$k^\alpha_{(2)} = 4c[(G - yG,_y)\cos(Dx/2) - (J - yJ,_y)\sin(Dx/2)]\delta^\alpha_0$$
$$+\{4c[G,_y\cos(Dx/2) - J,_y\sin(Dx/2)] - \frac{cd}{b^2 + ce}\}\delta^\alpha_1$$
$$+2cD[G\sin(Dx/2) + J\cos(Dx/2)]\delta^\alpha_2 + 4c[M\cos(Dx/2) - N\sin(Dx/2)]\delta^\alpha_3,$$

$$k^\alpha_{(3)} = \{-4b[(G - yG,_y)\cos(Dx/2) - (J - yJ,_y)\sin(Dx/2)]$$
$$-2D[(G - yG,_y)\sin(Dx/2) + (J - yJ,_y)\cos(Dx/2)]\}\delta^\alpha_0$$
$$+\{-4b[G,_y\cos(Dx/2) - J,_y\sin(Dx/2)]$$

$$-2D[G_{,y}\sin(Dx/2)+J_{,y}\cos(Dx/2)]+\frac{bd}{b^2+ce}\}\delta^\alpha_1$$
$$+\{2bD[-G\sin(Dx/2)-J\cos(Dx/2)]+D^2[G\cos(Dx/2)-J\sin(Dx/2)]\}\delta^\alpha_2$$
$$+\{-4b[M\cos(Dx/2)-N\sin(Dx/2)]-2D[M\sin(Dx/2)+N\cos(Dx/2)]\}\delta^\alpha_3. \qquad (5.12)$$

By changing the basis to $k'^\alpha_{(2)}=k^\alpha_{(2)}+(b^2+ce)^{-1}cdk^\alpha_{(1)}$ and $k'^\alpha_{(3)}=k^\alpha_{(3)}-(b^2+ce)^{-1}bdk^\alpha_{(1)}$ the result equivalent to:

$$d=h=0 \qquad (5.13)$$

is achieved, and then, from (5.8):

$$G^2+J^2=-(gy^2+2B)/(4cD^2). \qquad (5.14)$$

Eq. (5.14) suggests the parametrization:

$$G=K\cos L, \qquad J=K\sin L, \qquad (5.15)$$

then, from (5.14):

$$K^2=G^2+J^2=-(gy^2+2B)/(4cD^2), \qquad (5.16)$$

and $L$ remains arbitrary. With $a=d=0$ we change the basis again to $k"^\alpha_{(2)}=(4c)^{-1}k'^\alpha_{(2)}$ and $k"^\alpha_{(3)}=(-2D)^{-1}[k'^\alpha_{(3)}+(b/c)k'^\alpha_{(2)}]$ and we carry out the transformation (1.13) with:

$$H=2L/D, \qquad T_{,y}=-2(M\sin L-N\cos L)/(DK). \qquad (5.17)$$

The result is equivalent to $L=0=N$ which implies:

$$J=0, \qquad G=K, \qquad M=\gamma/(DK), \qquad (5.18)$$

and the Killing fields become:

$$k^\alpha_{(1)}=\delta^\alpha_1,$$

$$k^\alpha_{(2)}=(K-yK_{,y})\cos(Dx/2)\delta^\alpha_0+K_{,y}\cos(Dx/2)\delta^\alpha_1$$
$$+\frac{1}{2}DK\sin(Dx/2)\delta^\alpha_2+[\gamma/(DK)]\cos(Dx/2)\delta^\alpha_3,$$

$$k^\alpha_{(3)}=(K-yK_{,y})\sin(Dx/2)\delta^\alpha_0+K_{,y}\sin(Dx/2)\delta^\alpha_1$$
$$-\frac{1}{2}DK\cos(Dx/2)\delta^\alpha_2+[\gamma/(DK)]\sin(Dx/2)\delta^\alpha_3. \qquad (5.19)$$

The commutation relations are:

$$[k_1,k_2]=-\frac{1}{2}Dk_3, \qquad [k_2,k_3]=-[g/(8cD)]k_1, \qquad [k_3,k_1]=-\frac{1}{2}Dk_2 \qquad (5.20)$$

The Bianchi type depends on the constant $g$: for $g/c > 0$ it is type IX, for $g/c < 0$ it is type VIII and for $g = 0$ it is type $\mathrm{VII}_0$. The last case is contained in (4.20) - (4.23) as the subcase $b + f = 0$.

The Killing equations for $k^\alpha_{(1)}$ imply that the metric tensor is independent of $x$. Knowing this, one can simplify the Killing equations for $k^\alpha_{(2)}$ and $k^\alpha_{(3)}$. Since $k^\alpha_{(2)}$ and $k^\alpha_{(3)}$ are linear in $\sin(Dx/2)$ and $\cos(Dx/2)$, while $g_{\alpha\beta}$ are independent of $x$, each Killing equation implies two equations: the coefficients of $\cos(Dx/2)$ and of $\sin(Dx/2)$ have to vanish separately. The pair of equations implied by $k^\alpha_{(2)}$ is identical to the pair implied by $k^\alpha_{(3)}$, and it is:

$$\begin{aligned}(K - yK,_y)g_{\alpha\beta,t} + [\gamma/(DK)]g_{\alpha\beta,z} + \frac{1}{4}D^2K\delta^1_\alpha g_{2\beta} \\ +\delta^2_\alpha\{-yK,_{yy}\, g_{0\beta} + K,_{yy}\, g_{1\beta} - [\gamma K,_y/(DK^2)]g_{3\beta}\} + \frac{1}{4}D^2K\delta^1_\beta g_{\alpha 2} \\ +\delta^2_\beta\{-yK,_{yy}\, g_{\alpha 0} + K,_{yy}\, g_{\alpha 1} - [\gamma K,_y/(DK^2)]g_{\alpha 3}\} = 0, \end{aligned} \tag{5.21}$$

$$\begin{aligned}Kg_{\alpha\beta,y} + \delta^1_\alpha\{-(K - yK,_y)g_{0\beta} - K,_y\, g_{1\beta} - [\gamma/(DK)]g_{3\beta}\} \\ +K,_y\,\delta^2_\alpha g_{2\beta} + \delta^1_\beta\{-(K - yK,_y)g_{\alpha 0} - K,_y\, g_{\alpha 1} - [\gamma/(DK)]g_{\alpha 3}\} + K,_y\,\delta^2_\beta g_{\alpha 2} = 0,\end{aligned} \tag{5.22}$$

The solution of (5.22) is:

$$g_{11} = y^2 + 2H_{33}(\gamma K/D)^2 \int K^{-3}R(y)dy + 2(\gamma/D)K^2H_{13}R(y) + K^2H_{11},$$

$$g_{12} = (\gamma/D)H_{23}R(y) + H_{12}, \qquad g_{13} = (\gamma/D)H_{33}KR(y) + KH_{13},$$

$$g_{22} = H_{22}/K^2, \qquad g_{23} = H_{23}/K, \qquad g_{33} = H_{33}, \tag{5.23}$$

where the $H_{ij}(t, z)$ are arbitrary functions, and $R(y)$ is:

$$R(y) := \int K^{-3}\mathrm{d}y. \tag{5.24}$$

With $K$ given by (5.16), $R(y)$ and $\int K^{-3}R(y)dy$ can be easily calculated, but the result has to be given separately for $gB \neq 0$, for $g = 0$ and for $B = 0$, so eq. (5.24) is the most compact notation (but see below).

Note that $\gamma$ and $B$ cannot vanish simultaneously; if $\gamma = 0 = B$, then $K - yK,_y = 0 = K,_{yy}$, and then eq. (5.21) implies $g_{12} = g_{22} = g_{23} = 0$. Together with $g_{02} = 0$ (we are still in the Plebański class) this means that $\det(g_{\alpha\beta}) = 0$. Also, $g$ and $B$ cannot vanish simultaneously because with $g = 0 = B$ we are back in the domain of Paper 2. With $\gamma^2 + B^2 \neq 0$, the following new variables can be introduced for solving (5.21):

$$u = 2cD\gamma t + Bz, \qquad v = -Bt + 2cD\gamma z. \tag{5.25}$$

With $\gamma$ and $B$ running through all possible values, the hypersurfaces $u = \text{const}$ are timelike, null or spacelike. However, the solution of the Killing equations has the same dependence on $u$ and $v$ in every case.

For solving (5.21), the cases $gB \neq 0$ and $gB = 0$ have to be separated.

**Case I:** $gB \neq 0$.
(this means we are considering the Bianchi types IX and VIII here, but not $\mathrm{VII}_0$). In this case:

$$R = -2cD^2 y/(BK), \tag{5.26}$$

and the solution of (5.21) (with (5.23) - (5.24) already taken into account) is:

$$\lambda^2 := gB/(8\delta^4 D^2), \qquad \delta^2 := (B/D)^2 + (2c\gamma)^2, \qquad U := h_{12}\sinh(2\lambda v) + k_{12}\cosh(2\lambda v),$$

$$H_{11} = -cD^2 U/(2\delta^2\lambda) + h_{11}, \qquad H_{12} = h_{12}\cosh(2\lambda v) + k_{12}\sinh(2\lambda v),$$

$$H_{13} = -[cD^2/(2\delta^2\lambda)][h_{23}\sinh(\lambda v) + k_{23}\cosh(\lambda v)],$$

$$H_{22} = -2\delta^2\lambda U/(cD^2) - [gB/(2c^2D^6)]h_{11} - [8c\gamma^2/(BD^2)]h_{33},$$

$$H_{23} = h_{23}\cosh(\lambda v) + k_{23}\sinh(\lambda v), \qquad H_{33} = h_{33}, \tag{5.27}$$

where the $h_{ij}(u)$ and $k_{ij}(u)$ are arbitrary functions.

Eqs. (5.27) are adapted to the case $gB > 0$. When $gB < 0, \lambda^2 < 0$, i.e. $\lambda$ is imaginary. Then $k_{12}$ and $k_{23}$ have to be taken imaginary (and $U$ becomes imaginary in consequence of this), the hyperbolic functions go over into the corresponding trigonometric functions in the well-known way, and the functions $H_{ij}$ remain real.

**Case II:** $gB = 0$.

Then $K,_{yy} = 0$. The form of $R(y)$ still depends on whether $g \neq 0$ or $g = 0$ and $B \neq 0$ or $B = 0$. The formulae below apply in each case. The solution of (5.21) (again with (5.23) - (5.24) already taken into account) is here:

$$H_{11} = \frac{(c\gamma gy)^2}{64\delta^8 D^2 K^2}h_{33}v^4 - \frac{c^2\gamma D gy}{8\delta^6}h_{23}v^3 + (\frac{c^2D^4}{4\delta^4}h_{22} + \frac{c\gamma gy}{4\delta^4 DK}h_{13})v^2 - \frac{cD^2}{\delta^2}h_{12}v + h_{11},$$

$$H_{12} = -\frac{c(\gamma gy)^2}{16\delta^6 D^4 K^2}h_{33}v^3 + \frac{3c\gamma gy}{8\delta^4 D}h_{23}v^2 - (\frac{cD^2}{2\delta^2}h_{22} + \frac{\gamma gy}{2\delta^2 D^3 K}h_{13})v + h_{12},$$

$$H_{13} = \frac{c\gamma gy}{8\delta^4 DK}h_{33}v^2 - \frac{cD^2}{2\delta^2}Kh_{23}v + h_{13}, \qquad H_{22} = \frac{(\gamma gy)^2}{4\delta^4 D^6 K^2}h_{33}v^2 - \frac{\gamma gy}{\delta^2 D^3}h_{23}v + h_{22},$$

$$H_{23} = -\frac{\gamma gy}{2\delta^2 D^3 K}h_{33}v + Kh_{23}, \qquad H_{33} = h_{33}. \tag{5.28}$$

The subcase of (5.28) in which $g = 0$ and the hypersurfaces $u =$ const are spacelike should have a common subset witht the class considered by Demiański and Grishchuk[3]. These authors considered Bianchi type $\mathrm{VII}_0$ models with nonzero rotation, with spacelike orbits of the symmetry group which are flat and with the source being a perfect fluid (the pressure is not constant in their class). A member of the present collection should result when $p =$ const. However, Ref. 3 does not contain sufficient information to identify it.

With this, Case 1.1 is exhausted. We go back to (2.8) with $\Delta \neq 0$ and consider:

## VI. Case 1.2 : $\mathbf{c} = 0$.

Just as it was announced in the paragraph after (2.25), this case brings no new information. Three situations occur here:

1. The group becomes two-dimensional (because two Killing vectors become collinear); these cases are not considered here.

2. A linear combination of the Killing fields with constant coefficients is spanned on $u^\alpha$ and $w^\alpha$; these cases are in the domain of Paper 2.

3. In the case when the group is 3-dimensional and none of the Killing fields is spanned on $u^\alpha$ and $w^\alpha$, the formulae are equivalent to (2.25) - (2.29) and (3.2) - (3.11) (both sets reappear).

The proof consists simply in retracing the whole reasoning from (2.8) on with $c = 0$. As seen from (2.8), with $c = 0$ necessarily $\Delta \geq 0$, and so no analog of Case 1.1.2 arises here. The essential steps of the reasoning are described in Appendix B.

Case 1 is exhausted at this point.

**VII. Case 2: $\Delta = 0$ (i.e. $A$ has one double eigenvalue).**

The reasoning from (2.8) on has to be repeated with this new assumption. Then the double eigenvalue is $\alpha$ and:

$$b = f \pm 2\sqrt{-ce}, \qquad \alpha = f \pm \sqrt{-ce} = (b + f)/2. \tag{7.1}$$

The double sign denotes two different cases, but they will be considered at one go. In finding the solutions of (2.5), the cases $\det A \neq 0$ and $\det A = 0$ have to be considered separately. We first consider:

**Case 2.1:** $\det A \neq 0$.

This means:

$$b + f \neq 0 \neq \alpha. \tag{7.2}$$

In the next step, the case $c = 0$ has to be set aside for separate consideration, so we first assume:

**Case 2.1.1:** $c \neq 0$.

The solutions for $\lambda_{2,3}$, $\phi$ and $\psi$ are here:

$$\begin{aligned}\lambda_2 &= \{[1 + (b - f)x/2]L_2(y) + \ cxL_3(y)\}e^{(b+f)x/2},\\ \lambda_3 &= \{-(b - f)^2 xL_2(y)/(4c) + [1 + (f - b)x/2]L_3(y)\}e^{(b+f)x/2},\end{aligned} \tag{7.3}$$

$$\begin{aligned}\phi &= \{[1 + (b - f)x/2]F(y) + \ cxP(y)\}e^{(b+f)x/2} + 4(cd - af)y/(b + f)^2 - C_2,\\ \psi &= \{-(b - f)^2 xF(y)/(4c) + [1 + (f - b)x/2]P(y)\}e^{(b+f)x/2}\\ &\quad -[a(b - f)^2/c + 4bd]y/(b + f)^2 - C_3,\end{aligned} \tag{7.4}$$

With such $\lambda_{2,3}$, $\phi$ and $\psi$, eq. (2.4$c$) involves polynomials of second degree in $e^{(b+f)x/2}$ and also of second degree in $x$. The equations implied by the coefficients of $x^2e^{(b+f)x}$ and of $xe^{(b+f)x}$ are fulfilled identically, while the remaining ones are:

$$-(4c)^{-1}(b - f)^2 FF,_y + fF,_y P - \ bFP,_y - \ cPP,_y = 0, \tag{7.5}$$

$$[(b-f)F/2+cP][d-h+(a+j)(b-f)/(2c)]=0, \tag{7.6}$$
$$[(b-f)F/2+cP]\{[a(b-3f)+4cd]/(b+f)-j\}=0, \tag{7.7}$$
$$g=\{-4h(cd-af)+j[a(b-f)^2/c+4bd]\}/(b+f)^2. \tag{7.8}$$

The vanishing of the first factor in (7.6) and (7.7) leads to the relation:

$$k^\alpha_{(3)}+\frac{b-f}{2c}k^\alpha_{(2)}-\frac{a(f-b)-2cd}{c(b+f)}k^\alpha_{(1)}=(\frac{b-f}{2c}L_2+L_3)e^{(b+f)x/2}\delta^\alpha_3,$$

which means that the combination on the left is collinear with $w^\alpha$, and so this case belongs to the domain of Paper 2. Hence, (7.6) and (7.7) imply:

$$h=d+(a+j)(b-f)/(2c), \qquad j=[a(b-3f)+4cd]/(b+f). \tag{7.9}$$

The solution of (7.5) may be represented parametrically by:

$$F=2(\mathcal{A}R+cR\ln R)/(b+f), \qquad P=-(b-f)F/(2c)+R, \tag{7.10}$$

where $\mathcal{A}$ is an arbitrary constant and $R(y)$ is an arbitrary function. It can be assumed that $R\neq 0$ because otherwise we are back in the domain of Paper 2.

Eq. (2.3$l$), with the functions given by (7.3) - (7.4) is a polynomial of the same form as (2.4$c$), and only the coefficients of $e^{(b+f)x}$ provide a new equation:

$$-(4c)^{-1}(b-f)^2(F_{,y}L_2+FL_{2,y})+fF_{,y}L_3-bFL_{3,y}-c(L_{3,y}P+L_3P_{,y})$$
$$-bL_2P_{,y}+FL_{2,y}P=0. \tag{7.11}$$

Using (7.10), the solution is found again in a parametric form:

$$L_2=2S(\mathcal{A}+c\ln R+c)/(b+f)+\gamma R, \qquad L_3=-(b-f)L_2/(2c)+S, \tag{7.12}$$

where $\gamma$ is another arbitrary constant and $S$ is another arbitrary function. The resulting Killing fields are now calculated from (2.1) using (7.10) and (7.12). Since this procedure has already been performed a few times in this paper, we shall not quote the intermediate results. It turns out that $k^\alpha_{(2)}$ and $k^\alpha_{(3)}$ contain terms which are constant multiples of $k^\alpha_{(1)}$, these are removed when a new basis, $k'^\alpha_{(2)}$ and $k'^\alpha_{(3)}$ is appropriately defined. Then we change $k^\alpha_{(3)}$ again, to $k"^\alpha_{(3)}=k'^\alpha_{(3)}+[(b-f)/(2c)]k'^\alpha_{(2)}$ and carry out the transformation (1.13) with:

$$H=2\ln R/(b+f), \qquad T=2(b+f)^{-1}\int(S/R)\mathrm{d}y. \tag{7.13}$$

The Killing fields that result are (with primes dropped):

$$k^\alpha_{(1)}=\delta^\alpha_1,$$
$$k^\alpha_{(2)}=e^{(b+f)x/2}\{[cx+2\mathcal{A}/(b+f)]\delta^\alpha_0-[c+\mathcal{A}+(b+f)cx/2]\delta^\alpha_2+\gamma\delta^\alpha_3\}$$
$$k^\alpha_{(3)}=e^{(b+f)x/2}[\delta^\alpha_0-\frac{1}{2}(b+f)\delta^\alpha_2]. \tag{7.14}$$

It can be assumed now that $\mathcal{A} = 0$ because this is equivalent to changing the basis to $k'^{\alpha}_{(2)} = k^{\alpha}_{(2)} - 2\mathcal{A}k^{\alpha}_{(3)}/(b+f)$. The commutation relations are:

$$[k_{(1)}, k_{(2)}] = \frac{1}{2}(b+f)k_{(2)} + ck_{(3)}, \qquad [k_{(2)}, k_{(3)}] = 0, \qquad [k_{(3)}, k_{(1)}] = -\frac{1}{2}(b+f)k_{(3)}, \quad (7.15)$$

and they correspond to Bianchi type IV.

Because of $[k_{(2)}, k_{(3)}] = 0$, coordinates can be adapted to $k_{(2)}$ and $k_{(3)}$ simultaneously. The following transformation does it:

$$t' = e^{-(b+f)x/2}[t + \frac{1}{2}(b+f)tx + xy], \qquad x' = x,$$
$$y' = -c^{-1}e^{-(b+f)x/2}[\frac{1}{2}(b+f)t + y], \qquad z' = \gamma(b+f)t/(2c) + \gamma y/c + z, \quad (7.16)$$

and it results in (primes dropped):

$$k^{\alpha}_{(1)} = -[\frac{1}{2}(b+f)t + cy]\delta^{\alpha}_0 + \delta^{\alpha}_1 - \frac{1}{2}(b+f)y\delta^{\alpha}_2, \qquad k^{\alpha}_{(2)} = \delta^{\alpha}_2, \qquad k^{\alpha}_{(3)} = \delta^{\alpha}_0,$$
$$u^{\alpha} = e^{-(b+f)x/2}[(W/c)\delta^{\alpha}_0 - (2c)^{-1}(b+f)\delta^{\alpha}_2 + \gamma(2c)^{-1}(b+f)\delta^{\alpha}_3] \qquad w^{\alpha} = n\delta^{\alpha}_3. \quad (7.17)$$

In the new coordinates the metric is independent of $t$ and $y$. After the coordinate transformation (7.16) and after solving the Killing equations for $k^{\alpha}_{(1)}$ it assumes the form:

$$g_{00} = \frac{1}{4}(b+f)^2e^{(b+f)x}h_{22}, \qquad g_{01} = \frac{1}{2}(b+f)e^{(b+f)x/2}h_{12},$$
$$g_{02} = e^{(b+f)x}[-2c/(b+f) + \frac{1}{2}(b+f)Wh_{22} + \frac{1}{2}(b+f)\gamma h_{23}],$$
$$g_{03} = \frac{1}{2}(b+f)e^{(b+f)x/2}h_{23}, \qquad g_{11} = h_{11},$$
$$g_{12} = e^{(b+f)x/2}(Wh_{12} + \gamma h_{13}), \qquad g_{13} = h_{13},$$
$$g_{22} = e^{(b+f)x}\{-8c(b+f)^{-2}W + [2c/(b+f)]^2 + 2\gamma Wh_{23} + \gamma^2h_{33} + W^2h_{22}\},$$
$$g_{23} = e^{(b+f)x/2}(Wh_{23} + \gamma h_{33}), \qquad g_{33} = h_{33}, \quad (7.18)$$

where the $h_{ij}(z)$ are arbitrary functions, and $W$, not to be confused with the same symbol from sec. IV, is:

$$W := \frac{1}{2}(b+f)cx + \ c. \quad (7.19)$$

**VIII. Case 2.1.2:** $c = 0$.

It follows from (7.1) that in this case:

$$b = f = \alpha \neq 0. \quad (8.1)$$

Similarly as it happened in sec. VI, it turns out that this case is included in the case $c \neq 0$. When the procedure of sec. VII is retraced with (8.1), the Killing fields that result at the stage corresponding to (7.14) are equivalent to (7.14), with $b + f = 2b$, $e$ playing the role of $c$ and the roles of $k_{(2)}$ and $k_{(3)}$ interchanged.

**IX. Case 2.2:** $\det A = 0$.

With $\Delta = \det A = 0$, the following follows from (2.8) and (2.6):

$$bf = ce, \qquad f = -b, \qquad \alpha_1 = \alpha_2 = 0, \tag{9.1}$$

and the matrix $A$ is nilpotent, $A^2 = 0$. Again the case $c = 0$ has to be set aside for separate consideration, so we first follow:

**Case 2.2.1:** $c \neq 0$.

The solutions of (2.5) - (2.6) are here:

$$\lambda_2 = L_2(y) + \ cxL_3(y), \qquad \lambda_3 = -(b/c)\lambda_2 + L_3(y), \tag{9.2}$$

$$\phi = F(y) + \ cxP(y) + \frac{1}{2}(ab + cd)x^2 y,$$

$$\psi = -(b/c)\phi + P(y) + c^{-1}(ab + cd)xy + c^{-1}(-ay - bC_2 - cC_3). \tag{9.3}$$

In considering eqs. (2.4$c$) and (2.3$l$), the case $ab + cd = 0$ has to be considered separately, so now we follow:

**Case 2.2.1.1:** $ab + cd \neq 0$.

Eq. (2.4$c$) is now a polynomial of second degree in $x$, and it implies the following equations:

$$j = a, \qquad h = -d, \qquad (ab + cd)(yF_{,y} + F + C_2) - c^2 PP_{,y} = (cg - a^2)y. \tag{9.4}$$

The integral of the last equation is:

$$F = \frac{1}{2}(ab + cd)^{-1}[(cP)^2/y + (cg - a^2)y] + \mathcal{A}(ab + cd)/y - C_2. \tag{9.5}$$

Eq. (2.3$l$) reduces to the single equation:

$$-c^2(L_{3,y}P + L_3 P_{,y}) + (ab + cd)(yL_{2,y} + L_2) = 0. \tag{9.6}$$

In solving (9.6) the case $P = 0$ has to be considered separately. However, although some of the intermediate steps of the calculation depend on the assumption $P \neq 0$, in the final formulae for the Killing fields $P = 0$ is achieved by a transformation of the (1.13) set. The result is identical to the one obtained with $P = 0$ from (9.6) on.

With $P \neq 0$, the solution of (9.6) is:

$$L_3 = (ab + cd)[yL_2 - \gamma(ab + cd)]/(c^2 P). \tag{9.7}$$

We change the basis of the resulting Killing fields by taking $k'^{\alpha}_{(3)} = k^{\alpha}_{(3)} + (a/c)k^{\alpha}_{(1)} + (b/c)k^{\alpha}_{(2)}$, then carry out the transformation (1.13) with:

$$H = cP/[y(ab+cd)], \qquad T_{,y} = cL_3/[y(ab+cd)], \tag{9.8}$$

and again change the basis to $k'^{\alpha}_{(2)} = (ab+cd)^{-1}[k^{\alpha}_{(2)} - \frac{1}{2}(ab+cd)^{-1}(cg-a^2)k^{\alpha}_{(1)}]$, $k"^{\alpha}_{(3)} = c(ab+cd)^{-1}k'^{\alpha}_{(3)}$. The final Killing fields are:

$$k^{\alpha}_{(1)} = \delta^{\alpha}_1, \qquad k^{\alpha}_{(2)} = (2\mathcal{A}/y)\delta^{\alpha}_0 + (-\mathcal{A}/y^2 + x^2/2)\delta^{\alpha}_1 - xy\delta^{\alpha}_2 + (\gamma/y)\delta^{\alpha}_3,$$
$$k^{\alpha}_{(3)} = x\delta^{\alpha}_1 - y\delta^{\alpha}_2. \tag{9.9}$$

The commutation relations are:

$$[k_{(1)}, k_{(2)}] = k_{(3)}, \qquad [k_{(2)}, k_{(3)}] = -k_{(2)}, \qquad [k_{(3)}, k_{(1)}] = -k_{(1)}, \tag{9.10}$$

and they correspond to Bianchi type VIII.

The Killing equations for $k_{(1)}$ imply that the metric is independent of $x$. Those for $k_{(3)}$ are solved by:

$$g_{11} = H_{11}y^2, \qquad g_{12} = H_{12}, \qquad g_{13} = H_{13}y,$$
$$g_{22} = H_{22}/y^2, \qquad g_{23} = H_{23}/y, \qquad g_{33} = H_{33}, \tag{9.11}$$

where the $H_{ij}(t,z)$ are arbitrary functions (the components $g_{0\alpha}$ are as in (1.9), we are still in the Plebański class in this case). In solving the Killing equations for $k_{(2)}$ it is useful to observe that $\mathcal{A}$ and $\gamma$ cannot vanish simultaneously: with $\mathcal{A} = \gamma = 0$, the Killing equations for $k_{(2)}$ imply that $g_{12} = g_{22} = g_{23} = 0$, i.e. $\det(g_{\alpha\beta}) = 0$. With $\mathcal{A}^2 + \gamma^2 \neq 0$, the following variables can be introduced:

$$u = \gamma t - 2\mathcal{A}z, \qquad v = 2\mathcal{A}t + \gamma z. \tag{9.12}$$

The Killing equations for $k^{\alpha}_{(2)}$ have to be solved separately for $\mathcal{A} \neq 0$ and for $\mathcal{A} = 0$. When $\mathcal{A} \neq 0$, we define $\delta$ and $\lambda$ by:

$$\delta^2 = 4\mathcal{A}^2 + \gamma^2, \qquad \lambda^2 = -2\mathcal{A}/\delta^4, \tag{9.13}$$

and then the solution is:

$$U := h_{12}\sinh(2\lambda v) + k_{12}\cosh(2\lambda v), \qquad V := h_{23}\sinh(\lambda v) + k_{23}\cosh(\lambda v),$$
$$H_{11} = (-2\mathcal{A})^{-1/2}U + (\gamma/\mathcal{A})(-2\mathcal{A})^{-1/2}V + h_{11},$$
$$H_{12} = h_{12}\cosh(2\lambda v) + k_{12}\sinh(2\lambda v) + \gamma H_{23}/(2\mathcal{A}),$$
$$H_{13} = (-2\mathcal{A})^{-1/2}V + \gamma h_{33}/(2\mathcal{A}), \qquad H_{22} = (-2\mathcal{A})^{1/2}U + 2\mathcal{A}h_{11} - \gamma^2 h_{33}/(2\mathcal{A}),$$
$$H_{23} = h_{23}\cosh(\lambda v) + k_{23}\sinh(\lambda v), \qquad H_{33} = h_{33}, \tag{9.14}$$

where the $h_{ij}(u)$ are arbitrary functions. Eqs. (9.14) are adapted to the case $\lambda^2 > 0$ ($\mathcal{A} < 0$), but the solution for $\lambda^2 < 0$ can be easily constructed from this one.

When $\mathcal{A} = 0$, necessarily $\gamma \neq 0$ and the Killing equations for $k^{\alpha}_{(2)}$ are solved in the original variables $(t, z)$ as follows:

$$H_{11} = h_{33}z^4/(2\gamma)^2 + h_{23}z^3/\gamma^2 + (h_{22}/\gamma^2 + h_{13}/\gamma)z^2 + 2h_{12}z/\gamma + h_{11},$$

$$H_{12} = h_{33}z^3/(2\gamma) + 3h_{23}z^2/(2\gamma) + (h_{22}/\gamma + h_{13})z + h_{12},$$
$$H_{13} = h_{33}z^2/(2\gamma) + h_{23}z/\gamma + h_{13}, \qquad H_{22} = h_{33}z^2 + 2h_{23}z + h_{22},$$
$$H_{23} = h_{33}z + h_{23}, \qquad H_{33} = h_{33}, \tag{9.15}$$

where the $h_{ij}(t)$ are arbitrary functions.

Eqs. (9.14) and (9.15) are very similar in form to (5.27) and (5.28), respectively. This suggests that Case 2.2.1.1 considered here may be included in Case 1.1.2.2 of sec. V as a limit (combined with a coordinate transformation). However, this author was not able to prove or disprove this hypothesis.

We go back now to (9.3) and consider:

**X. Case 2.2.1.2:** $ab + cd = 0$.

Eqs. (9.2) still apply and eqs. (9.3) simplify in the obvious way. The difference with sec. IX is that eqs. (2.4$c$) and (2.3$l$) impose weaker conditions here. The case $P = 0$ leads to the domain of Paper 2, so it may be assumed that $P \neq 0$. Then, the consequences of (2.4$c$) are:

$$h = bj/c, \tag{10.1}$$
$$cPP,_y + (j - a)P + (g - aj/c)y = 0. \tag{10.2}$$

Eq. (10.2) defines $P$ as a function of $y$. The consequence of eq. (2.3$l$) is:

$$c(L_3P),_y + (j - a)L_3 = 0, \tag{10.3}$$

which defines $L_3$ once $P$ is given. The functions $F$ and $L_2$ are still arbitrary at this point.

In the resulting Killing fields we take $k'^{\alpha}_{(3)} = k^{\alpha}_{(3)} + (a/c)k^{\alpha}_{(1)} + (b/c)k^{\alpha}_{(2)}$ instead of $k^{\alpha}_{(3)}$, and carry out the transformation (1.13) with:

$$H = (F + C_2)/(cP), \qquad T,_y = (L_2/c - HL_3)/P. \tag{10.4}$$

The result of the change of basis and of (10.4) is equivalent to:

$$F = -C_2, \qquad L_2 = a = b = d = 0, \qquad c = 1, \tag{10.5}$$

and the Killing fields become:

$$k^{\alpha}_{(1)} = \delta^{\alpha}_1, \qquad k^{\alpha}_{(2)} = x(P - yP,_y)\delta^{\alpha}_0 + xP,_y \delta^{\alpha}_1 - P\delta^{\alpha}_2 + xL_3\delta^{\alpha}_3,$$
$$k^{\alpha}_{(3)} = (P - yP,_y)\delta^{\alpha}_0 + P,_y \delta^{\alpha}_1 + L_3\delta^{\alpha}_3. \tag{10.6}$$

Note that with (10.5) fulfilled, the subcase $g = 0$ implies $P,_y = \text{const}$, and then $k^{\alpha}_{(3)} - P,_y k^{\alpha}_{(1)}$ is spanned on $u^{\alpha}$ and $w^{\alpha}$, i.e. this case is in the domain of Paper 2. Hence, in what follows it will be assumed that $g \neq 0$. For the same reason it will also be assumed that $(P - yP,_y) \neq 0$.

The commutation relations (with (10.5) already taken into account) are:

$$[k_{(1)}, k_{(2)}] = k_{(3)}, \qquad [k_{(2)}, k_{(3)}] = jk_{(3)} + gk_{(1)}, \qquad [k_{(3)}, k_{(1)}] = 0. \tag{10.7}$$

The collection of Bianchi types contained in (10.7) is as follows. When:

(a) $g < j^2/4$, the Bianchi type is $\mathrm{VI}_h$ with the free parameter $a_B = j/(j^2 - 4g)^{1/2}$; and $\mathrm{VI}_0$ when $j = 0$.

(b) $g > j^2/4$, the Bianchi type is $\mathrm{VII}_h$ with the free parameter $a_B = j/(4g - j^2)^{1/2}$; and $\mathrm{VII}_0$ when $j = 0$.

(c) $g = j^2/4$, the Bianchi type is IV if $j \neq 0$ and II if $j = 0$. This last case belongs to the domain of Paper 2, as explained above, and will not be presented here.

(d) $g = 0$ with no condition on $j$, the Bianchi type is III.

The Killing equations have to be solved separately for these three cases.

The coordinates are adapted to $k_{(1)}$ and $k_{(3)}$ by the following transformation:

$$\begin{aligned} t' = t/(P - yP,_y), \qquad x' = x - tP,_y/(P - yP,_y), \qquad y' = y, \\ z' = -L_3 t/(P - yP,_y) + z \end{aligned} \tag{10.8}$$

In the new coordinates, with primes dropped:

$$\begin{aligned} k_{(1)}^\alpha = \delta_1^\alpha, \qquad k_{(2)}^\alpha = (x - jt)\delta_0^\alpha - gt\delta_1^\alpha - P\delta_2^\alpha, \qquad k_{(3)}^\alpha = \delta_0^\alpha, \\ u^\alpha = (P - yP,_y)^{-1}(\delta_0^\alpha - P,_y\delta_1^\alpha - L_3\delta_3^\alpha), \qquad w^\alpha = n\delta_3^\alpha, \\ g_{00} = P^2 - y^2 {P,_y}^2 + {P,_y}^2 g_{11} + 2L_3 P,_y g_{13} + {L_3}^2 g_{33}, \\ g_{01} = y(P - yP,_y) + P,_y g_{11} + L_3 g_{13}, \\ g_{02} = P,_y g_{12} + L_3 g_{23}, \qquad g_{03} = P,_y g_{13} + L_3 g_{33}. \end{aligned} \tag{10.9}$$

The Killing equations for $k_{(1)}$ and $k_{(3)}$ imply now that the metric is independent of $t$ and $x$, while those for $k_{(2)}$ are:

$$\begin{aligned} -Pg_{11,y} + 2y(P - yP,_y) + 2P,_y g_{11} + 2L_3 g_{13} = 0, \\ -Pg_{12,y} + L_3 g_{23} = 0, \qquad -Pg_{13,y} + P,_y g_{13} + L_3 g_{33} = 0, \\ -Pg_{22,y} - 2P,_y g_{22} = 0, \qquad -Pg_{23,y} - P,_y g_{23} = 0, \qquad -Pg_{33,y} = 0. \end{aligned} \tag{10.10}$$

The last three have the same solutions in all three cases:

$$g_{22} = h_{22}/P^2, \qquad g_{23} = h_{23}/P, \qquad g_{33} = h_{33}, \tag{10.11}$$

where the $h_{ij}(z)$ in the formulae (10.11) - (10.21) are arbitrary functions. The remaining equations in (10.10) have to be solved separately for each case. In each case it is useful to introduce the new variable $Y(y)$ by:

$$Y,_y = 1/P. \tag{10.12}$$

Then eq. (10.3) (with (10.5) taken into account) has the solution:

$$L_3 = (\gamma/P)e^{-jY}. \tag{10.13}$$

In order to find $P$ one has to differentiate (10.2) (again with (10.5) taken into account) by $Y$ and use $y,_Y = P$. The equation becomes:

$$P_{,YY} + jP_{,Y} + gP = 0, \tag{10.14}$$

and $y$ is then found from $y = \int PdY$. The solutions of (10.10) and of (10.14) are now as follows:

**Case I:** $g < j^2/4$.

$$P = M\mathrm{e}^{\mu_1 Y} + Ne^{\mu_2 Y}, \qquad y = (M/\mu_1)e^{\mu_1 Y} + (N/\mu_2)e^{\mu_2 Y}, \tag{10.15}$$

where $M$ and $N$ are arbitrary constants and:

$$\mu_{1,2} = -j/2 + \varepsilon_{1,2}(j^2/4 - g)^{1/2}, \qquad \varepsilon_1 = 1, \varepsilon_2 = -1. \tag{10.16}$$

It can be assumed that $MN \neq 0$ because in both the cases $M = 0$ and $N = 0$ we are back in the domain of Paper 2 (then $P_{,y} = \text{const}$). The solutions of the remaining Killing equations are here:

$$g_{11} = y^2 + h_{11}P^2 - 2\gamma h_{13}e^{\mu_2 Y}P/[M(\mu_1 - \mu_2)] + \gamma^2 h_{33}e^{2\mu_2 Y}/[M(\mu_1 - \mu_2)]^2,$$

$$g_{12} = h_{12} - \gamma h_{23}e^{\mu_2 Y}/[M(\mu_1 - \mu_2)P], \qquad g_{13} = h_{13}P - \gamma h_{33}e^{\mu_2 Y}/[M(\mu_1 - \mu_2)]. \tag{10.17}$$

**Case II:** $g > j^2/4$.

Here we define:

$$D = (g - j^2/4)^{1/2}, \qquad U := M\cos(DY) + N\sin(DY),$$
$$V := M\sin(DY) - N\cos(DY), \tag{10.18}$$

where $M$ and $N$ are arbitrary constants, and then the solutions are:

$$P = e^{-jY/2}U, \qquad y = (4D^2 + j^2)^{-1}e^{-jY/2}(4DV - 2jU),$$
$$g_{11} = y^2 + h_{11}P^2 + 2\gamma h_{13}e^{-jY}UV/[D(M^2 + N^2)] + \gamma^2 h_{33}e^{-jY}/[D^2(M^2 + N^2)],$$
$$g_{12} = h_{12} + \gamma h_{23}V/[D(M^2 + N^2)U], \qquad g_{13} = h_{13}P + \gamma h_{33}PV/[D(M^2 + N^2)U]. \tag{10.19}$$

**Case III:** $g = j^2/4 \neq 0$.

With $M$ and $N$ being arbitrary constants, the solutions for $P$ and $y$ are here:

$$P = (MY + N)e^{-jY/2}, \qquad y = -2P/j - 4Me^{-jY/2}/j^2. \tag{10.20}$$

We can assume $M \neq 0$ because with $M = 0$ again $P_{,y} = \text{const}$. Then, the solutions of the Killing equations are as follows:

$$g_{11} = y^2 + h_{11}P^2 - 2\gamma h_{13}e^{-jY/2}P/M + \gamma^2 h_{33}e^{-jY}/M^2,$$
$$g_{12} = h_{12} - \gamma h_{23}/[M(My + N)], \qquad g_{13} = h_{13}P - \gamma h_{33}e^{-jY/2}/M. \tag{10.21}$$

**Case IV:** $g = 0$, no condition on $j$.

The conclusion that (10.5) can be achieved by a linear transformation of the basis of Killing fields is still valid. With $a = g = 0$ and $c = 1$, eqs. (10.2) and (10.3) are solved by:

$$P = -jy + M, \qquad L_3 = \gamma, \tag{10.22}$$

and the Killing fields are then:

$$k_{(1)}^{\alpha} = \delta_1^{\alpha}, \qquad k_{(2)}^{\alpha} = Mx\delta_0^{\alpha} - jx\delta_1^{\alpha} + (jy - M)\delta_2^{\alpha} + \gamma x\delta_3^{\alpha}, \qquad k_{(3)}^{\alpha} = M\delta_0^{\alpha} - j\delta_1^{\alpha} + \gamma\delta_3^{\alpha}. \tag{10.23}$$

The metric that is a solution of the Killing equations for the above is:

$$g_{11} = (2M/j)y - (M/j)^2 + (jy - M)^2 h_{11} + 2(\gamma/j)(jy - M)h_{13} + (\gamma/j)^2 h_{33},$$

$$g_{12} = h_{12} + \frac{\gamma h_{23}}{j(jy - M)}, \qquad g_{13} = (jy - M)h_{13} + (\gamma/j)h_{33},$$

$$g_{22} = h_{22}/(jy - M)^2, \qquad g_{23} = h_{23}/(jy - M), \qquad g_{33} = h_{33}, \tag{10.24}$$

where the $h_{ij}$ are arbitrary functions of the argument

$$T = \gamma t - Mz. \tag{10.25}$$

With this, case 2.2.1 is exhausted. We go back to (9.1) and consider:

**XI. Case 2.2.2 : c** $= 0$.

With $\Delta = \det A = c = 0$ the only element of the matrix $A$ that may be nonzero is $e$. The solutions of (2.5) - (2.6) are here:

$$\lambda_2 = L_2(y), \qquad \lambda_3 = exL_2 + L_3(y), \tag{11.1}$$

$$\phi = axy + F(y), \qquad \psi = \frac{1}{2}aex^2y + (dy + eC_2 + eF)x + P(y). \tag{11.2}$$

Eq. (2.4$c$) now implies that either $j = a$ or $a = 0$ or $e = 0$. However, $j = a$ leads to a result equivalent to (9.9) and $a = 0$ leads to a result equivalent to (10.6). Hence:

$$e = 0, \tag{11.3}$$

which makes the whole matrix $A = 0$. The remaining implications of (2.4$c$) and (2.3$l$) are:

$$ah + dj = 0, \qquad y(dF_{,y} - aP_{,y} - g) = h(F + C_2) + j(P + C_3),$$

$$y(dL_{2,y} - aL_{3,y}) = hL_2 + jL_3. \tag{11.4}$$

In solving these conditions, again several cases have to be considered separately.

**Case 2.2.2.1:** $a \neq 0$.

Then:

$$h = -dj/a, \tag{11.5}$$

but in solving the second of (11.4) the case $j = -a$ has to be set aside for separate consideration. We first follow:

**Case 2.2.2.1.1:** $j \neq -a$.
Then:

$$P = -gy/(j+a) + \; By^{-j/a} - C_3 + (d/a)(F + C_2),$$
$$L_3 = -\mathcal{A}y^{-j/a} + (d/a)L_2, \tag{11.6}$$

where $\mathcal{A}$ and $B$ are arbitrary constants. In the resulting Killing fields we change the basis to $k'^{\alpha}_{(2)} = a^{-1}k^{\alpha}_{(2)}$, $k'^{\alpha}_{(3)} = k^{\alpha}_{(3)} - (d/a)k^{\alpha}_{(2)} + [g/(j+a)]k^{\alpha}_{(1)}$, and then carry out the transformation (1.13) with:

$$H = (C_2 + F)/(ay), \qquad T_{,y} = L_2/(ay). \tag{11.7}$$

The final Killing fields are:

$$k^{\alpha}_{(1)} = \delta^{\alpha}_1, \qquad k^{\alpha}_{(2)} = x\delta^{\alpha}_1 - y\delta^{\alpha}_2,$$
$$k^{\alpha}_{(3)} = y^{-j/a}[B(j/a+1)\delta^{\alpha}_0 - Bj(ay)^{-1}\delta^{\alpha}_1 - \mathcal{A}\delta^{\alpha}_3]. \tag{11.8}$$

The commutation relations are:

$$[k_{(1)}, k_{(2)}] = k_{(1)}, \qquad [k_{(2)}, k_{(3)}] = (j/a)k_{(3)}, \qquad [k_{(3)}, k_{(1)}] = 0, \tag{11.9}$$

and they correspond to Bianchi type $\mathrm{VI}_h$ with the free parameter $a_B = (j-a)/(j+a)$.

The coordinates will be adapted to $k_{(1)}$ and $k_{(3)}$ after the transformation:

$$t' = aty^{j/a}/[B(j+a)], \qquad x' = jt/[(j+a)y] + x, \qquad y' = y, \qquad z' = \mathcal{A}at + B(j+a)z. \tag{11.10}$$

After the transformation:

$$k^{\alpha}_{(1)} = \delta^{\alpha}_1, \qquad k^{\alpha}_{(2)} = -jt\delta^{\alpha}_0 + ax\delta^{\alpha}_1 - ay\delta^{\alpha}_2, \qquad k^{\alpha}_{(3)} = \delta^{\alpha}_0,$$
$$u^{\alpha} = (j+a)^{-1}[aB^{-1}y^{j/a}\delta^{\alpha}_0 + (j/y)\delta^{\alpha}_1] + \mathcal{A}a\delta^{\alpha}_3, \qquad w^{\alpha} = B(j+a)n\delta^{\alpha}_3,$$
$$g_{00} = (By^{-j/a})^2[1 - (j/a)^2 + (j/a)^2 g_{11}/y^2 + 2\mathcal{A}j(j/a+1)g_{13}/y + \mathcal{A}^2(j+a)^2 g_{33}],$$
$$g_{01} = By^{-j/a}[(j/a+1)y - (j/a)g_{11}/y - \mathcal{A}(j+a)g_{13}],$$
$$g_{02} = -By^{-j/a}[(j/a)g_{12}/y + \mathcal{A}(j+a)g_{23}], \qquad g_{03} = -By^{-j/a}[(j/a)g_{13}/y + \mathcal{A}(j+a)g_{33}], \tag{11.11}$$

and the Killing equations imply:

$$g_{11} = h_{11}y^2, \qquad g_{12} = h_{12}, \qquad g_{13} = h_{13}y,$$
$$g_{22} = h_{22}/y^2, \qquad g_{23} = h_{23}/y, \qquad g_{33} = h_{33}, \tag{11.12}$$

where the $h_{ij}(z)$ are arbitrary functions.

**Case 2.2.2.1.2:** $j = -a$.

The formulae for $h$ and $L_3$ simply follow from (11.5) and (11.6) while the second of (11.4) now has a different integral:

$$P = (-g/a)y \ln y + \ By \ - C_3 + (d/a)(F + C_2). \tag{11.13}$$

In the resulting Killing fields we change the basis to $k'^{\alpha}_{(2)} = a^{-1}k^{\alpha}_{(2)}$, $k'^{\alpha}_{(3)} = k^{\alpha}_{(3)} - (d/a)k^{\alpha}_{(2)} + (g/a - B)k^{\alpha}_{(1)}$, and then carry out the transformation (1.13) with the same $H$ and $T$ as in (11.7). The final Killing fields $k^{\alpha}_{(1)}$ and $k^{\alpha}_{(2)}$ are the same as in (11.8), and:

$$k^{\alpha}_{(3)} = y\delta^{\alpha}_0 - \ln y \delta^{\alpha}_1 - \mathcal{A}y\delta^{\alpha}_3. \tag{11.14}$$

The commutation relations are:

$$[k_{(1)}, k_{(2)}] = k_{(1)}, \qquad [k_{(2)}, k_{(3)}] = k_{(1)} - k_{(3)}, \qquad [k_{(3)}, k_{(1)}] = 0, \tag{11.15}$$

and they correspond to Bianchi type IV.

In order to adapt the coordinates to $k^{\alpha}_{(1)}$ and $k^{\alpha}_{(2)}$ we now carry out the transformation:

$$t' = t/y, \qquad x' = (t/y)\ln y + x, \qquad y' = y, \qquad z' = \mathcal{A}t + z, \tag{11.16}$$

and it leads to:

$$\begin{gathered} k^{\alpha}_{(1)} = \delta^{\alpha}_1, \qquad k^{\alpha}_{(2)} = t\delta^{\alpha}_0 + (x - t)\delta^{\alpha}_1 - y\delta^{\alpha}_2, \qquad k^{\alpha}_{(3)} = \delta^{\alpha}_0, \\ u^{\alpha} = y^{-1}(\delta^{\alpha}_0 + \ln y \delta^{\alpha}_1) + \mathcal{A}\delta^{\alpha}_3, \qquad w^{\alpha} = n\delta^{\alpha}_3, \end{gathered}$$

$$\begin{gathered} g_{00} = y^2 - 2y^2 \ln y + (\ln y)^2 g_{11} + 2\mathcal{A}y \ln y g_{13} + (\mathcal{A}y)^2 g_{33}, \\ g_{01} = y^2 - \ln y g_{11} - \mathcal{A}y g_{13}, \qquad g_{02} = -\ln y g_{12} - \mathcal{A}y g_{23}, \\ g_{03} = -\ln y g_{13} - \mathcal{A}y g_{33}, \end{gathered} \tag{11.17}$$

and the Killing equations lead to formulae for the other metric components that are identical to (11.12).

**Case 2.2.2.2:** $a = 0$.

Then the first of (11.4) implies that either $d = 0$ or $j = 0$. However, if $d = 0$, then either the algebra of the Killing vectors necessarily becomes two-dimensional (this case is not considered here) or

$$j = 0 \tag{11.18}$$

follows anyway. Hence only (11.18) will be considered further. Then, with $0 \neq d \neq h$, a result equivalent to (11.8) follows, and when $0 \neq d = h$, a result equivalent to Case 2.2.2.1.2 follows. Therefore, new results are contained only in the case:

$$d = 0. \tag{11.19}$$

Then, from (11.4):

$$h(F + C_2) + gy = 0. \tag{11.20}$$

If $h \neq 0$, then (11.20) implies that the combination of the Killing fields $k_{(2)}^\alpha + (g/h)k_{(1)}^\alpha$ is collinear with $w^\alpha$, and this case is in the domain of Paper 2. Hence, from (11.20):

$$h = g = 0, \tag{11.21}$$

which leaves us with an Abelian algebra (Bianchi type I) and the functions $F$, $P$, $L_2$ and $L_3$ being all arbitrary. To avoid landing in the domain of Paper 2, we have to assume:

$$F,_y \neq 0 \neq P_y, \qquad C_2 + F - yF,_y \neq 0 \neq C_3 + P - yP,_y\,. \tag{11.22}$$

We will denote:

$$W_F := C_2 + F - yF,_y\,, \qquad W_P := C_3 + P - yP,_y\,, \tag{11.23}$$

and then the Killing fields are:

$$k_{(1)}^\alpha = \delta_1^\alpha, \qquad k_{(2)}^\alpha = W_F\delta_0^\alpha + F,_y\,\delta_1^\alpha + L_2\delta_3^\alpha, \qquad k_{(3)}^\alpha = W_P\delta_0^\alpha + P,_y\,\delta_1^\alpha + L_3\delta_3^\alpha. \tag{11.24}$$

It may also be assumed that:

$$\mathcal{L} := L_3 - L_2 W_P/W_F \neq 0 \tag{11.25}$$

because if $\mathcal{L} = 0$, then the Killing equations imply that either the symmetry group becomes two-dimensional or the metric is singular. With $\mathcal{L} \neq 0$, the following transformation is permissible:

$$t' = (L_3 t - W_P z)/(\mathcal{L}W_F), \qquad x' = (L_2 P,_y - L_3 F,_y)t/(\mathcal{L}W_F) + x - (P,_y - F,_y\,W_P/W_F)z/\mathcal{L},$$

$$y' = y, \qquad z' = -L_2 t/(\mathcal{L}W_F) + z/\mathcal{L}, \tag{11.26}$$

after which the Killing fields become:

$$k_{(1)}^\alpha = \delta_1^\alpha, \qquad k_{(2)}^\alpha = \delta_0^\alpha, \qquad k_{(3)}^\alpha = \delta_3^\alpha. \tag{11.27}$$

The resulting metric depends only on $y$, but the Einstein equations are hopelessly complicated.

This is the end of the collection of solutions of the Killing equations.

## XII. Perspectives.

The research for this series of papers was motivated by the desire to find a rotating (exact) perturbation of the Friedmann - Lemaître cosmological models. Several papers have been published whose authors found exact solutions of Einstein's equations with a rotating source (see a brief overview in sec. XIII), but all except one of them are either stationary from the beginning or become static in the limit of zero rotation. The one exception is the solution 2 by Stephani[4] that has still nonzero expansion in the limit $\omega \to 0$, but it is a rotating perturbation of a degenerate limit of the hyperbolically symmetric Kantowski -

Sachs solution[5] (see Ref. 6 for more on this point) and cannot reproduce any Robertson - Walker geometry.

There is no proof available that a rotating perturbation of the F-L models can be spatially homogeneous. However, if it has any spatially homogeneous subcases and has a dust source, then the subcases must be contained among the metrics listed in this series of papers - just because this is a complete list of all hypersurface-homogeneous rotating dust metrics (and, over all three papers, all Bianchi types appeared on the list, some of them more than once). Hence, from now on, instead of testing various metric ansatzes for rotating dust by trial and error, one can choose an ansatz from a limited collection. It is well-known[7] that the Robertson - Walker geometries have the following relation to the spatially homogeneous Bianchi-type geometries:

The spatially flat ($k = 0$) R-W geometry is a common subset of Bianchi types I and $\mathrm{VII}_0$.

The R-W geometry with negative spatial curvature ($k = -1$) is a common subset of Bianchi types V and $\mathrm{VII}_h$.

The R-W geometry with positive spatial curvature ($k = +1$) is a subset of Bianchi type IX geometries.

With this information, the following can be concluded:

1. The metrics from Paper 1 do not contain any generalization of the F-L models; the Bianchi type I class contained there has timelike symmetry orbits and the velocity field is tangent to the orbits.

2. The same is true for the Bianchi type I metrics from Paper 2 (cases 2.1.2.2 and 2.2) and from the present paper (eqs. (11.22) - (11.27)). This is in agreement with Theorem 3.1 of King and Ellis[8] which says that no tilted Bianchi type I models exist (tilted means that the velocity field is not orthogonal to the symmetry orbits. Bianchi models with rotation obviously must be tilted).

3. The Bianchi type V metrics from Paper 2 (cases 1.2.1 and 1.2.2.2) with suitably chosen parameters do have spacelike symmetry orbits and so may harbour some generalization of the $k = -1$ F-L model.

4. The Bianchi type $\mathrm{VII}_h$ metrics of the present paper (eqs. (4.23) and (10.19)) may have spacelike symmetry orbits at least on some open subsets of the manifold, i.e. they may contain generalizations of the $k = -1$ F-L model.

5. The same is true for the Bianchi type IX and type $\mathrm{VII}_0$ metrics of the present paper (eqs. (5.23) with (5.27) or (5.28)); generalizations of the $k = +1$ and $k = 0$ F-L models may be contained there.

Hence, the cases listed in points 3, 4 and 5 are most promising from the point of view of cosmology.

## XIII. A brief overview of literature.

Partly in order to justify the claim made in the first paragraph of sec. XII, a brief overview of literature on solutions of Einstein's equations with rotating sources will be presented here. For the period up to 1973, the overview is based on a thorough survey of subject indexes to *Physics Abstracts* starting with the year 1915 (made in connection with Refs. 57 - 59). For the period 1973 - 1996 the survey was less thorough - I was looking for only four keywords: "Bianchi", "homogeneous", "rotation" and "spatially homogeneous". In both searches the following sections of *Physics Abstracts* were surveyed: Cosmology, General Relativity, Gravitation and Space-Time Configurations (of course, references in the papers included in the survey were also checked). Ref. 9 is a more extended survey that covers the period up to 1973, and papers in which perturbative methods were used were also listed in it, these are omitted here. Vacuum solutions are omitted, too.

As a rule, each paper is mentioned in one sentence, so this overview does not pretend to represent the contents of the papers; it is only meant to sort the papers by subjects and serve as an introductory guide through the literature.

Lanczos[10] found the historically earliest exact solution with rotating matter (although he may have been unaware of its rotation); it is a dust solution in which the velocity field and the rotation field are collinear with Killing fields. Van Stockum[11] rediscovered the $\Lambda = 0$ subcase of the Lanczos solution; Ref. 11 contains in addition important contributions to the techniques of solving the Einstein equations for stationary axisymmetric metrics (i.e. with a two-dimensional symmetry group) with a rigigly rotating source. Ref. 12 is another rediscovery of the Lanczos solution, and Refs. 13 and 14 contain discussions of its properties.

Gödel's solution[15] has a five-dimensional symmetry group whose orbit is the whole spacetime, and consequently the physical scalars in it are all constant. Its source is dust of zero expansion and shear, and constant matter density and rotation scalars. Rediscoveries of the Gödel solution were published in Refs. 16 and 17 (see Ref. 18). Ref. 17 contains in addition a stationary axisymmetric perfect fluid solution.

Refs. 19 - 32 deal with properties of a metric form that is a modest generalization of the Gödel solution and is known in the literature as the "Gödel-type metric". This notion was introduced in Ref. 19, and in Ref. 20 (which in fact preceded Ref. 19) it was shown that the only Gödel-type metric with a perfect fluid source is the Gödel solution itself. However, various Gödel-type solutions with nonperfect fluid sources have been derived and investigated in Refs. 19 and 21 - 32.

In Refs. 33 - 40 solutions of Einstein - Maxwell equations with a charged fluid source were discussed, some of them under the additional assumption that the Lorentz force is zero. The latter are fully within the scope of the formalism used in the present series of papers because the charged dust in them moves with zero acceleration. The relation of the results of these papers to those obtained here was described in Paper 1. Some of them are generalizations of the Lanczos[10] and Gödel[15] solutions. Those from Refs. 33 - 39 are stationary cylindrically symmetric, i.e. are closely related to those from Paper 1, the one from Ref. 40 has a two-dimensional symmetry group. The solutions in Ref. 39 are coordinate transforms of those from Ref. 38.

A generalization of the $\Lambda = 0$ subcase of the Lanczos solution to a mixture of scalar field and dust was found by Santos and Mondaini[41].

Other generalizations of the Gödel solution were found in Refs. 42 - 50. Raval and Vaidya[42] found two solutions with anisotropic pressure, one of them nonstationary. Bray[43] found a collection of solutions with the rotating fluid immersed in a magnetic field. Novello and Rebouças[44], Kitamura[45−47] and Ray[48] found generalizations with heat flow (the second of Kitamura's solutions presented in Ref. 45 is a coordinate transform of the Gödel solution). Rebouças[49] found a generalization with free electromagnetic field; see Ref. 51 and Paper 1 for an explanation of the relation between the solutions in Refs. 49 and 37. Finally, Panov[50] found a generalization with two scalar fields, anisotropic fluid, null radiation and heat conduction.

Other solutions with rotating charged matter source were obtained in Refs. 52 - 55. The Einstein - Maxwell equations for stationary axisymmetric charged dust were analysed by Bonnor[56].

In addition, several solutions with a perfect fluid source (i.e. with nonconstant pressure) and with the same Bianchi type I symmetry as was considered in Paper 1 have been published. These include a family of solutions by this author[57−60] in which the velocity and the rotation fields are collinear with the Killing fields. For the solutions considered in Refs. 57 and 59 the proportionality factor between $w^\alpha$ and the Killing field is explicitly given. For the family of metrics from Ref. 60, the factor is an arbitrary function and the family is defined by a differential equation. The solutions by Davidson[61−62] are explicit examples from the family of Ref. 60, possibly they are also coordinate transforms of members of the family from Ref. 57 defined by certain fixed values of parameters, but Refs. 61 and 62 do not contain sufficient information for precise identification.

Nilsson and Uggla[63] did a qualitative analysis (using the theory of dynamical systems) of perfect fluid solutions with the same symmetry that obey the linear barotropic equation of state.

The list above includes papers that are related to Paper 1 of this series. Results related to those of Paper 2 are contained in Refs. 64 - 77.

Among the solutions found by Ellis[64] there are some that directly belong to the collection of Paper 2, they are identified and described in Paper 2. Also within the scope of Paper 2 are the results of King[65] who investigated properties of the subcase $\beta = h_{13} = 0$ of Case 2.1.2.2 (of Paper 2) and provided a few examples of explicit solutions. Other explicit solutions in King's class were found by Maitra[66], Zimmerman[67] (this reference was not given in Paper 2) and Vishveshwara and Winicour[68].

In Refs. 69 - 73 rotating dust solutions with four-dimensional symmetry groups were found; at least some of them have 3-dimensional subgroups and are within the domain of Paper 2. However, as mentioned in Paper 2, these papers do not contain sufficient information for a complete identification of all such subcases. The first three of the six solutions given in Ref. 70 are among the Case 2.1.2.2 metrics of Paper 2.

Davidson[74] found an example of solution with King's subclass of symmetry (stationary, cylindrically symmetric, differentially rotating) and with a perfect fluid source obeying the linear barotropic equation of state. Two other stationary cylindrically symmetric perfect fluid solutions were found by Garcia and Kramer[75], the first is differentially rotating and has the symmetry of King's subclass, the other is rigidly rotating and so has the Bianchi type I symmetry of the class considered in Paper 1. Nilsson and Uggla[76] analysed by the method of dynamical systems the Einstein equations for a metric with a Bianchi type II symmetry and a perfect fluid source that obeys the linear barotropic equation of state. Its set of Killing fields is in the subcase $\lambda_3 = 0$ of Case 1.1.2.2 of Paper 2. Stewart and Ellis[77] considered perfect fluid generalizations of the Ellis (dust) solutions from Ref. 64, they also considered sources with anisotropic pressure, viscosity and electric charge.

The only paper directly relevant to the present Paper 3 is Ref. 3, see sec. V. The remaining part of the present section is a list of papers in which various problems connected with rotating matter were discussed, but which are more remotely related to the present series of papers.

Stephani[4] found a solution that is unique in one more respect in addition to that mentioned in sec. XII: it is so far the only rotating matter (dust) solution with no symmetry. (Note: solution 3 in Ref. 4 is not a perfect fluid solution, contrary to the paper's statement).

Stationary axisymmetric (i.e. with two-dimensional symmetry groups) perfect fluid solutions were found in Refs. 78 - 83. Wahlquist[78] found a rigidly rotating solution with the equation of state $\epsilon = -3p + \mathrm{const}$ that, with specific values of two parameters, can describe the interior of a body with compact outer surface. Herlt[79] found a source of the NUT vacuum solution and Kramer[80] found another solution, both are rigidly rotating (Kramer's solution was rediscovered by Patra and Roy[81], see Ref. 82). Another class of rigidly rotating stationary axisymmetric solutions was found by Herlt[83].

Several stationary axisymmetric metrics were devised as nonperfect fluid sources of the Kerr solution. Results in this class that were published up to 1976 are reviewed in Ref. 84. Later, a few more papers on this subject were published, but in those that are known to the present author the source is either a surface distribution of matter or an energy-momentum tensor that does not correspond to any identifiable kind of matter, hence they are not mentioned here.

A spatially homogeneous solution of Bianchi type $\mathrm{VI}_0$ with a rotating perfect fluid source was found by Rosquist[85]; it has the equation of state $\epsilon = 3p$ and nonzero expansion. Spatially homogeneous solutions with heat-conducting sources were found in Refs. 86 - 89. Two other hypersurface-homogeneous perfect fluid solutions were found by Wainwright[90], one has a 3-dimensional symmetry group of unidentified Bianchi type, the other has a 4-dimensional multiply transitive symmetry group.

The remaining papers (Refs. 91 - 116) contain results obtained without finding explicit solutions of Einstein's equations. This part of the survey is likely to be incomplete.

Narlikar[91] proposed a metric form for a rotating dust model in which dust particles move on 3-cylinders (axial symmetry is not assumed) whose orthogonal sections are 2-dimensional surfaces of constant curvature. Winicour[98] reduced the Einstein equations with a stationary axisymmetric dust source to a sequence of integrations. This author[101] presented results of partial integration of the Einstein equations for a cylindrically symmetric nonstationary perfect fluid.

Bampi and Cianci[102] investigated spacetimes with an Abelian 2-dimensional group of symmetries that has null orbits; the example of exact solution provided is a vacuum. Wils and van den Bergh[104] showed that a stationary axisymmetric differentially rotating charged dust has either a nonvanishing Lorentz force of a nonconstant ratio of charge density to mass density.

In Refs. 105 - 110 general properties of rotating spatially homogeneous Bianchi type IX models were investigated without attempting to solve the Einstein equations. Of these, Ref. 109 gives a kinetic theory description of sources in such models and Ref. 110 gives a qualitative analysis of rotating mixmaster models.

Refs. 111 - 113 contain general considerations about rotating matter models, and Refs. 8 and 114 contain more general results, applicable also to nonrotating models, but having consequences for rotating models as well. Collins[111] considered properties of shearfree rotating perfect fluids. Again Collins[112] reviewed arguments for the hypothesis that shear being zero implies that either rotation or expansion is zero (apart from several specific examples, this is still unproven). Mason and Pooe[113] investigated properties of the Lie derivative of the rotation vector along the velocity vector in rigidly rotating matter. King and Ellis[8] investigated properties of tilted spatially homogeneous models and Nilsson and Uggla[114] investigated hypersurface homogeneous and hypersurface self-similar perfect fluid models by the method of dynamical systems.

Hawking[115] and Collins and Hawking[116] investigated limits set on the rotation parameter in Bianchi-type models by the observations of the CMB radiation.

Finally, there is a paper in the domain of the history of science. Ellis[117] described the influence of Gödel's ideas presented in Refs. 15 and 105 on the development of several concepts and research programs in relativity, such as, among other things, the Bianchi-type models, singularity theorems and causal structure of spacetime.

**Acknowledgement** Calculations for this paper were done with use of the algebraic computer program Ortocartan[118−119].

**Note added in proof.** The "Gödel-type metric of Refs. 19 - 32 was discussed also in Ref. 120. Other rotating solutions of Einstein - Maxwell equations were discussed in Refs. 121 - 122. I am grateful to M. J. Rebouças and A. Georgiou for this information.

## APPENDIX A

The coordinate transformation (4.21) changes the metric tensor from the Plebański form (1.9) to one in which the following relations hold (primes dropped, all the components displayed are expressed in the new coordinates):

$$\begin{aligned} g_{00} = {} & e^{(b+f)x}[\sin^2(Dx/2) - (V/W)^2\cos^2(Dx/2)] + (V/W)^2 g_{22} \\ & +2(V/W)\gamma e^{(b+f)x/2}[\cos(Dx/2) + (V/W)\sin(Dx/2)]g_{23} \\ & +\gamma^2 e^{(b+f)x}[\cos(Dx/2) + (V/W)\sin(Dx/2)]^2 g_{33}, \end{aligned}$$

$$g_{01} = (V/W)g_{12} + \gamma e^{(b+f)x/2}[\cos(Dx/2) + (V/W)\sin(Dx/2)]g_{13},$$

$$\begin{aligned} g_{02} = {} & e^{(b+f)x}\cos(Dx/2)[\sin(Dx/2) - (V/W)\cos(Dx/2)] + (V/W)g_{22} \\ & +\gamma e^{(b+f)x/2}[\cos(Dx/2) + (V/W)\sin(Dx/2)]g_{23}, \end{aligned}$$

$$g_{03} = (V/W)g_{23} + \gamma e^{(b+f)x/2}[\cos(Dx/2) + (V/W)\sin(Dx/2)]g_{33}.$$

## APPENDIX B
### The equivalence of case 1.2 and case 1.1.1.

These are the main points of the reasoning from (2.8) on in the case $c = 0$. When $c = 0$, eqs. (2.7) - (2.8) imply:

$$\alpha_1 = b, \qquad \alpha_2 = f, \tag{B.1}$$

and $b \neq f$ since $\Delta \neq 0$. The calculation requires checking several cases separately.

**Case 1.2.1:** $\det A \neq 0$.

The formulae corresponding to (2.9) and (2.13) are then:

$$\lambda_2 = L_2(y)e^{bx}, \qquad \lambda_3 = [e/(b-f)]L_2(y)e^{bx} + L_3(y)e^{fx}, \qquad \phi = F(y)e^{bx} - ay/b - C_2,$$
$$\psi = [e/(b-f)]F(y)e^{bx} + P(y)e^{fx} - [-ae/(bf) + d/f]y - C_3. \tag{B.2}$$

In verifying (2.4$c$) and (2.3$l$) the cases $F = 0$ and $P = 0$ have to be considered separately. The results are as follows:

When $F = 0 \neq P$ and $L_2 = 0$ the group becomes 2-dimensional.

When $F = 0 \neq P$ and $L_2 \neq 0$, the Killing field $[k^\alpha_{(2)} + (a/b)k^\alpha_{(1)}]$ is collinear with $w^\alpha$, and this case is in the domain of Paper 2.

When $P = 0$, another linear combination of the Killing vectors with constant coefficients is collinear with $w^\alpha$.

When $F \neq 0 \neq P$, formulae equivalent to (2.24) - (2.29) result.

**Case 1.2.2:** $\det A = 0$.

Two different (but equivalent) subcases have to be considered here: $b = 0$ and $f = 0$ ($b \neq f$ because of the assumption $\Delta \neq 0$).

When $f = 0$, the functions $\lambda_2, \lambda_3$ and $\phi$ are given by the limit $f = 0$ of (B.2), while $\psi$ is:

$$\psi = (e/b)F(y)e^{bx} + P(y) - aey/b^2 + (d - ae/b)xy - C_3. \tag{B.3}$$

Eqs. (2.4$c$) and (2.3$l$) show then that there always exists a linear combination of the Killing fields with constant coefficients which is spanned on $u^\alpha$ and $w^\alpha$.

When $b = 0$, the conclusion is the same, only the functions in (B.2) are different at the starting point. $\lambda_2$ and $\lambda_3$ are the limits $b = 0$ of those from (B.2), and:

$$\phi = F(y) + axy - C_2,$$
$$\psi = -(e/f)(\phi + C_2) + P(y)e^{fx} - (d/f + ae/f^2)y - C_3. \tag{B.4}$$

## CAPTION TO THE DIAGRAM

The classes of metrics considered in the paper. Arrows point from more general classes to subclasses. The numbers at arrows are the case-numbers used in the text. The first entry in each rectangle is the property defining the case; all the symbols are introduced in eqs. (2.1) - (2.8). The subsequent entries give the following information: 1. The Bianchi type of the corresponding algebra (2.2); 2. The equation-numbers corresponding to the final result in the given case. No progress was made with the Einstein equations in any of the cases. Apart from case 1.1.2.2 for which a subcase was discussed in Ref. 3, none of the cases seem to have appeared in earlier literature (see sec. XIII).

The diagram does not show the links to the entries in the corresponding diagram in Paper 2; they are numerous and would obscure the drawing.

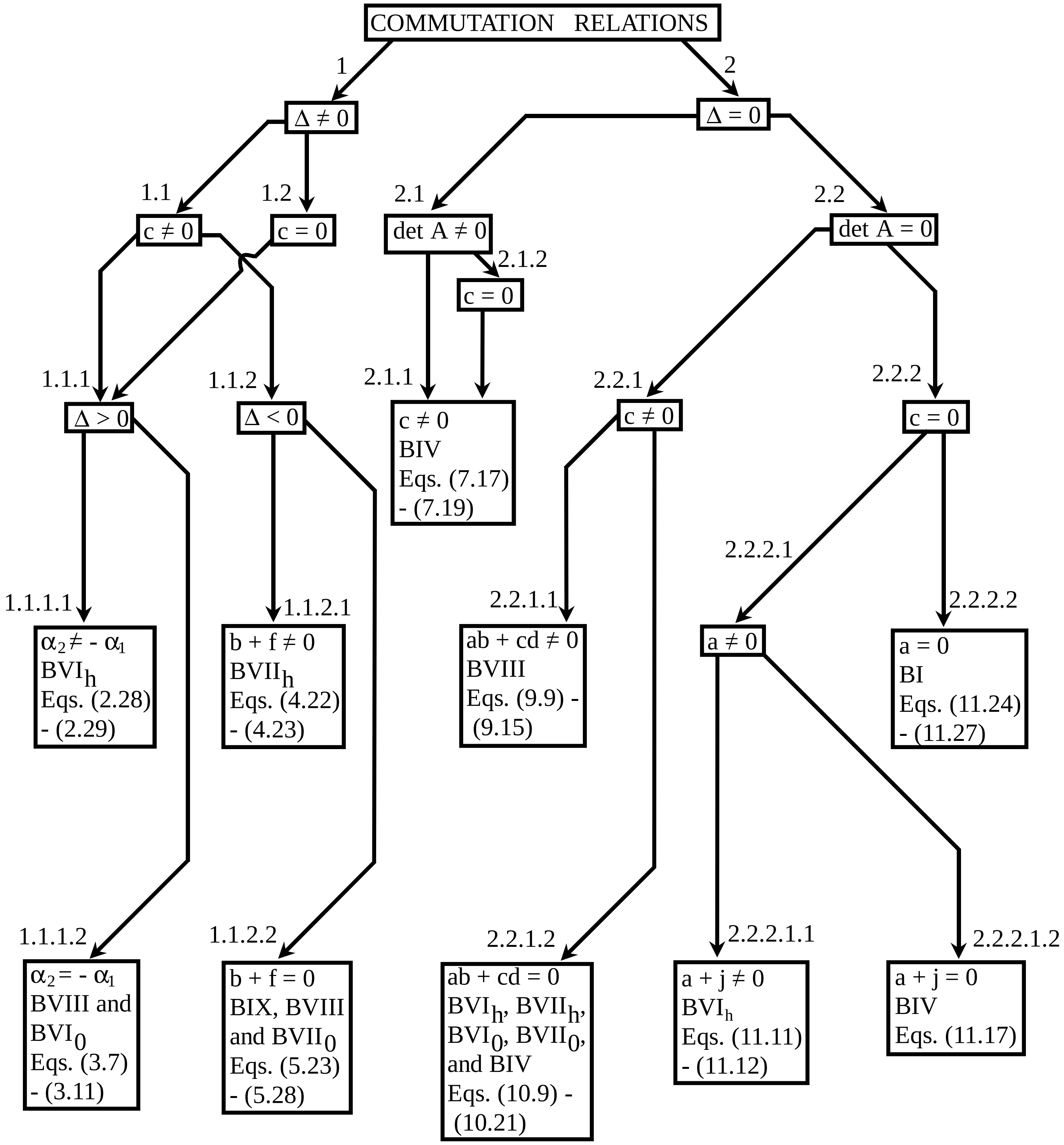

COMMUTATION RELATIONS
1
2
Δ ≠ 0
Δ = 0
1.1
1.2
2.1
2.2
c ≠ 0
c = 0
det A ≠ 0
det A = 0
2.1.2
c = 0
1.1.1
1.1.2
2.1.1
2.2.1
2.2.2
Δ > 0
Δ < 0
c ≠ 0 BIV Eqs. (7.17) - (7.19)
c ≠ 0
c = 0
2.2.2.1
1.1.1.1
1.1.2.1
2.2.1.1
2.2.2.2
$\alpha_2 \neq - \alpha_1$ $BVI_h$ Eqs. (2.28) - (2.29)
b + f ≠ 0 $BVII_h$ Eqs. (4.22) - (4.23)
ab + cd ≠ 0 BVIII Eqs. (9.9) - (9.15)
a ≠ 0
a = 0 BI Eqs. (11.24) - (11.27)
1.1.1.2
1.1.2.2
2.2.1.2
2.2.2.1.1
2.2.2.1.2
$\alpha_2 = - \alpha_1$ BVIII and $BVI_0$ Eqs. (3.7) - (3.11)
b + f = 0 BIX, BVIII and $BVII_0$ Eqs. (5.23) - (5.28)
ab + cd = 0 $BVI_h$, $BVII_h$, $BVI_0$, $BVII_0$, and BIV Eqs. (10.9) - (10.21)
a + j ≠ 0 $BVI_h$ Eqs. (11.11) - (11.12)
a + j = 0 BIV Eqs. (11.17)